\documentclass[pdflatex,sn-mathphys]{sn-jnl}% Math and Physical Sciences 
\usepackage{fix-cm}
\jyear{2025}%

\theoremstyle{thmstyleone}%
\theoremstyle{thmstyletwo}%

\theoremstyle{thmstylethree}%
\usepackage{xcolor, soul}
\sethlcolor{green}
\usepackage{lmodern}
\usepackage{anyfontsize}
\newcommand{\units}{\,\mathrm} % to print nice units within mathmode
\usepackage{comment}

\def\gev{\hbox{Ge\kern-.08em V\/}}

\usepackage{lineno}
\title{Novel Silicon and GaAs Sensors for Compact Sampling Calorimeters}

\author[1]{\fnm{H.} \sur{Abramowicz}}
\author[2]{\fnm{M.} \sur{Almanza Soto}}
\author[1]{\fnm{Y.} \sur{Benhammou}}
\author[1]{\fnm{M.} \sur{Elad}}
\author[3]{\fnm{M.} \sur{Firlej}}
\author[3]{\fnm{T.} \sur{Fiutowski}}
\author[4]{\fnm{V.} \sur{Ghenescu}}
\author[5]{\fnm{G.} \sur{Grzelak}}
\author[1]{\fnm{D.} \sur{Horn}}
\author[1,2]{\fnm{S.} \sur{Huang}}
\author[3]{\fnm{M.} \sur{Idzik}}
\author[2]{\fnm{A.} \sur{Irles}}
\author[1]{\fnm{A.} \sur{Levy}}
\author[1,6]{\fnm{I.} \sur{Levy}}
\author*[7,8]{\fnm{W.} \sur{Lohmann}}\email{wolfgang.lohmann@desy.de}
\author[3]{\fnm{J.} \sur{Moro\'{n}}}
\author[4]{\fnm{A. T.} \sur{Neagu}}
\author[3]{\fnm{D.} \sur{Pietruch}}
\author[4]{\fnm{P.M.} \sur{Potlog}}
\author[3]{\fnm{K.} \sur{\'Swientek}}
\author[5]{\fnm{A.F.} \sur{Żarnecki}}
\author[5]{\fnm{K.} \sur{Zembaczyński}}

\affil[1]{\orgdiv{School of Physics and Astronomy}, \orgname{Tel Aviv University}, \orgaddress{
    \city{Tel Aviv-Yafo} \postcode{69978}, \country{Israel}}}
\affil[2]{\orgdiv{IFIC}, \orgname{CSIC and Universitat de Val\`encia}, \orgaddress{
    \street{C/ Catedr\`atic Jos\'e Beltr\'an Mart\'inez 2},
    \postcode{46980} \city{Paterna}, \country{Spain}}}
\affil[3]{\orgdiv{Faculty of Physics and Applied Computer Science}, \orgname{AGH University of Krakow}, \orgaddress{
    \postcode{30-059} \city{Krak\'{o}w}, \country{Poland}}}
\affil[4]{\orgname{Institute of Space Science}, \orgaddress{
    \postcode{077125} \city{Bucharest}, \country{Romania}}}
\affil[5]{\orgdiv{Faculty of Physics}, \orgname{University of Warsaw}, \orgaddress{
    \street{Pasteura 5},
    \postcode{02-093} \city{Warszawa}, \country{Poland}}}
\affil[6]{\orgdiv{Department of Physics}, \orgname{Nuclear Research Centre-Negev}, \orgaddress{P.O. Box 9001}, \city{Beer Sheva}, \country{Israel}}   
\affil[7]{\orgname{Deutsches Elektronen Synchrotron (DESY)}, \orgaddress{
    \street{Platanenallee 6},
    \postcode{15738} \city{Zeuthen}, \country{Germany}}}
\affil[8]{\orgdiv{Institute of Physics}, \orgname{Brandenburg University of Technology}, \orgaddress{
    \street{Platz der Deutschen Einheit 1},
    \postcode{03046} \city{Cottbus},  \country{Germany}}}
\abstract{
Two samples of silicon pad sensors and two samples of GaAs sensors are studied in an electron beam with 5\,GeV energy
from the DESY-II test-beam facility.
The sizes of the silicon and GaAs sensors are about 9$\times$9\,cm$^2$ and 5$\times$8\,cm$^2$, respectively. The thickness is 500\,\textmu m for both the silicon and GaAs sensors. The pad size is about 5$\times$5\,mm$^2$. 
The sensors are foreseen to be used in a compact electromagnetic sampling calorimeter. 
The readout of the pads is done via traces connected to the pads and the front-end ASICs at the edges of the sensors.
For the silicon sensors, copper traces on a Kapton foil are connected to the sensor pads with conducting glue.
The pads of the GaAs sensors are connected to bond-pads via aluminium traces on the sensor substrate.
The readout is based on a dedicated front-end ASIC, called FLAME.
Pre-processing of the raw data and deconvolution is performed with FPGAs.
The whole system is orchestrated by a Trigger Logic Unit.
Results are shown for the signal-to-noise ratio, the homogeneity of the response, edge effects on pads, cross talk and wrongly assigned signals due to the readout traces.
}


\begin{document}
\maketitle

\newpage

\section{Introduction}\label{sec1}
For several applications of electromagnetic calorimeters, the Moli\`{e}re radius~\cite{Moliere_1, Moliere_2} is an important parameter. 
It is a measure of the transversal spread of a shower originating from an electron or photon in matter. It is proportional to the radiation length X$_0$. In materials of low X$_0$ the shower is squeezed inside a small radius, facilitating its detection in case of background, and ensuring a precise shower position measurement. In addition, the probability to resolve overlapping showers is enhanced. In practice, what matters is the effective X$_0$ which depends on the absorber and the space used for signal extraction. For compact calorimetry, the latter has to be minimized.  

Examples of relevance are luminometers in experiments at electron-positron colliders~\cite{Abramowicz:2010bg} or an electromagnetic calorimeter in the laser-electron scattering experiment LUXE~\cite{LUXE:2023crk} investigating strong field QED. In the former, Bhabha scattering~\cite{Bhabha:1936zz} is used as a gauge process. Using a highly compact calorimeter, i.e. with a small Moli\`{e}re radius, the fiducial volume is well defined, and the space needed is relatively small. In addition, the measurement of the shower of a high energy electron on top of widely spread low energy background is improved. In the 
laser-electron scattering case, the number of secondary electrons and positrons per bunch crossing varies over a wide range, and both the determination of the number of electrons and positrons and their energy spectrum per bunch crossing favours a highly compact calorimeter.
In sampling calorimeters, tungsten is the favoured absorber material, with a Moli\`{e}re radius of about 9.3\,mm. Tungsten plates are interspersed with active pad sensors to form a sandwich. To keep the Moli\`{e}re radius near the one of tungsten, the gap between tungsten plates must be kept small. Hence, thin sensor-planes are needed. 
In this case, solid state sensors are favoured because the thickness can be kept to a few hundreds micrometers while the generated signals are sufficiently large to detect even minimum ionising particles with high efficiency.  

Here two technologies of ultra-thin sensor planes are investigated based on silicon (Si) and gallium arsenide (GaAs) pad sensors\footnote{Compared to Si pad sensors GaAs sensors show a larger radiation tolerance under electron exposure at room temperature~\cite{Afanaciev:2012im,Kruchonak:2020jsk}.}. In both, thin metal traces guide the signal on a pad to the sensor edge where the front-end ASICs are positioned. For the GaAs sensors, these traces are made of aluminium, 
embedded in the gaps between the pads, and for the Si case, Kapton fan-outs with copper traces are glued to the sensor. 
Both technologies are new. Traces embedded between pads of the GaAs sensor are used for the first time. A similar technology was developed previously 
to improve the granularity of silicon strip sensors~\cite{deBoer:2015dmy}.
Kapton fan-outs with copper traces bonded on the sensor pads are described in a previous publication~\cite{Abramowicz:2018vwb}. Here electrically conductive glue is used to connect the copper traces to the pads, requiring less space than wire bonds. Using these connectivity schemes the thickness of the sensor plane can be easily kept below 1\,$\units{mm}$. 

In the LUXE experiment two electromagnetic calorimeters are foreseen to measure the multiplicity and energy spectrum of electrons and positrons produced in the multi-photon Breit-Wheeler process~\cite{Breit:1934zz}. The design, prototyping and construction of these calorimeters is pursued by the LUXE ECAL team. In this paper a full system test of two considered  pad-structured sensors for the LUXE ECAL is reported. The sensors are positioned in a $5\units{GeV}$ electron beam at DESY. A pixel telescope is used to measure the trajectory of each triggered electron. The signal on each pad is amplified and digitised by a dedicated front-end electronics. Results are presented on the signal-to-noise ratio, the homogeneity of the response of different pads, edge effects, cross talk and signals due to the readout traces on the GaAs substrate.

\section{Sensors}\label{sec2}

\begin{figure}[h!]
 \begin{center}
    \includegraphics[width=0.7\columnwidth]{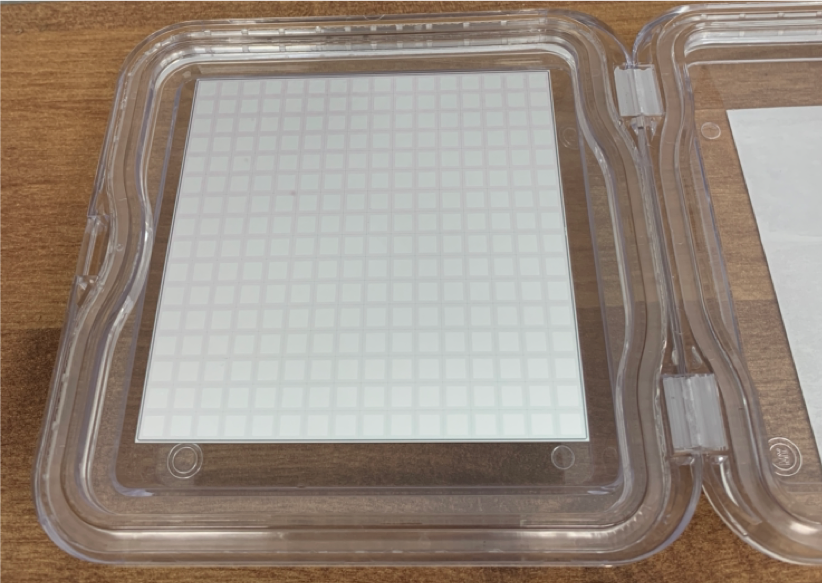}
    \caption{Picture of the Si sensor with $16 \times 16$ pads, each pad with a $5.52 \times 5.52 \units{mm^2}$ aluminium metallisation, in its protection box.}
    \label{fig:sensor_silicon}
 \end{center}
\end{figure}

Silicon sensors, produced by Hamamatsu initially for the CALICE collaboration~\cite{tomita2014study}, are arrays of $5.52\times 5.52 \units{mm^2}$, $p+$ on $n$ substrate diodes. The thickness is 500 \textmu m and the resistivity is $3 \units{k\Omega\ cm}$.
Each sensor has a total area of $89.9 \times 89.9 \units{mm^2}$, structured in $16 \times 16$ pads covered with aluminium, without guard rings. The gap between pads is 10 \textmu m. The active area of the sensor is surrounded by a 0.61$\units{mm}$ wide inactive zone. A picture of a sensor is shown in Fig.~\ref{fig:sensor_silicon}.
\begin{figure}[h!]
   \includegraphics[width=0.99\columnwidth]{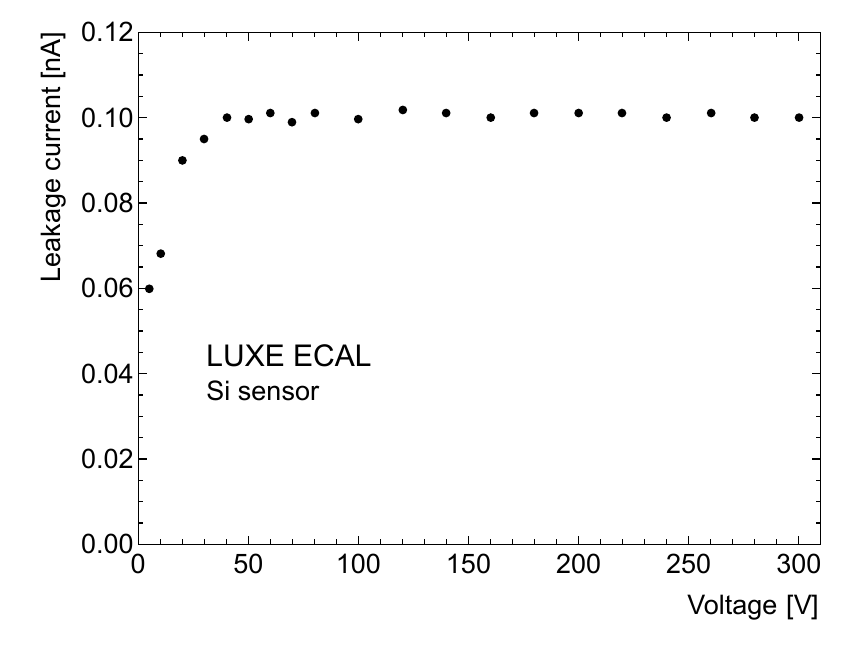}
    \caption{Leakage current as a function of the applied voltage for a selected pad of the Si sensor measured at $20^\circ$C.}
    \label{fig:sensor_silicon_IV}
\end{figure}

For all pads of the two sensors studied, the leakage current was measured as a function of the bias voltage. A typical result is shown in Fig.~\ref{fig:sensor_silicon_IV}.
Full depletion of the sensor is reached at a bias voltage of about $50 \units{V}$.

\begin{figure}[h!]
\begin{center}
    \includegraphics[width=0.65\columnwidth]{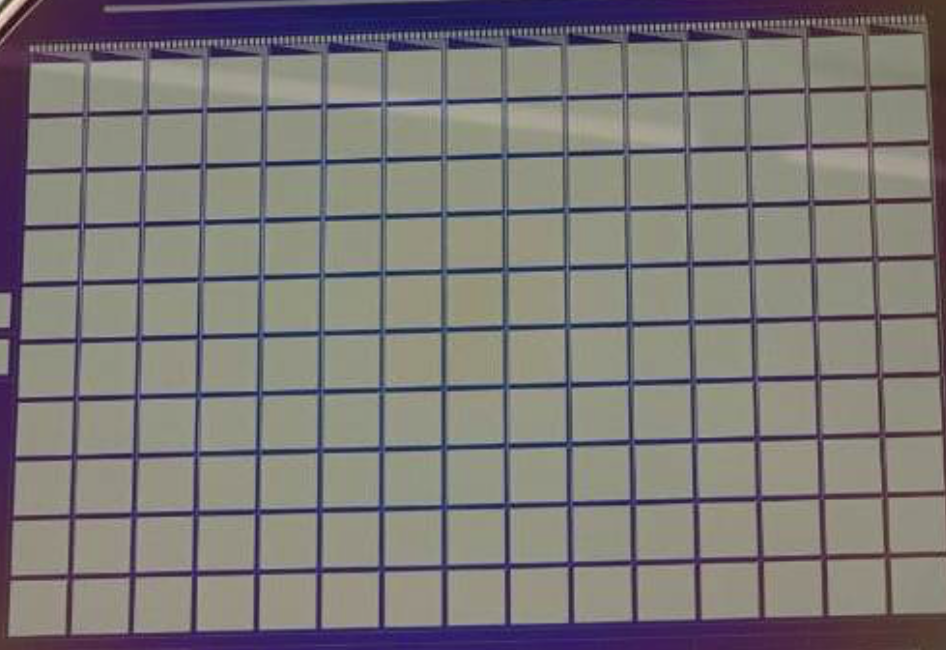}
    \caption{Picture of a GaAs sensor. The bond-pads are visible on top of the sensor.}
    \label{sensor_GAAS_pic}
\end{center}
\end{figure}

\begin{figure}[ht!]
\begin{center}
    \includegraphics[width=0.85\columnwidth]{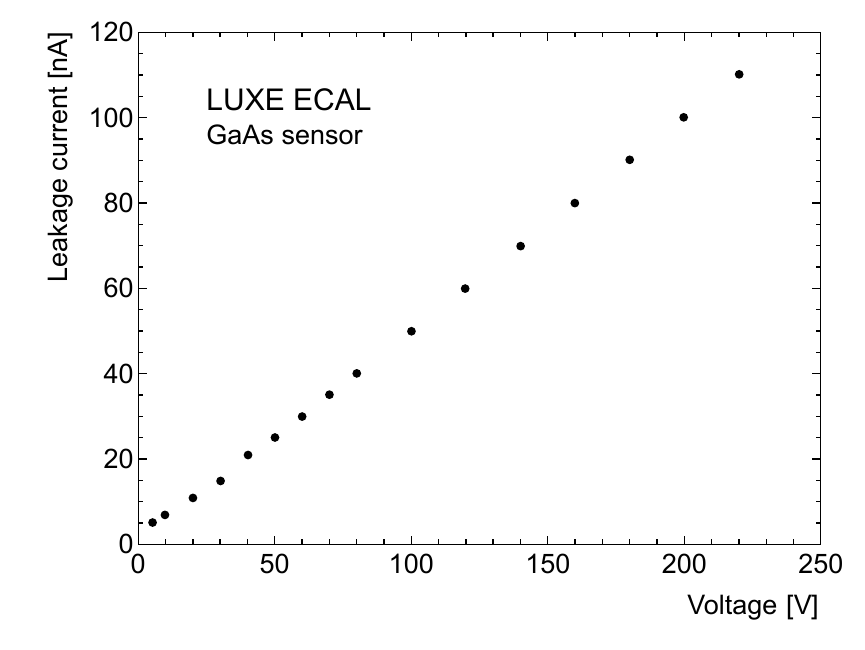}
    \caption{The leakage current of a GaAs pad as a function of the bias voltage, measured at $20^\circ$C.}
    \label{sensor_GaAs_leakage}
\end{center}
\end{figure}

GaAs sensors~\cite{tyazhev2021} are made of single crystals. High resistivity of $10^6 \units{k\Omega\ cm}$ is reached by compensation with chromium. The pads are $4.7 \times 4.7 \units{mm^2}$, with $0.3 \units{mm}$ gap between pads. Pads consist of a 0.05 \textmu m vanadium layer, covered with 1 \textmu m aluminium, made with electron beam evaporation and magnetron sputtering. The back-plane is made of nickel and aluminium of $0.02$ and 1 \textmu m thickness, respectively. 
The sensors are 500 \textmu m thick with an active area of $49.7 \times 74.7 \units{mm^2}$ leading to $10\times 15$ pads without guard rings. 
The signals from the pads are routed to bond pads on the top edge of the sensor      
by aluminium traces embedded in the gap between pads, thus avoiding the presence of a flexible printed circuit board (PCB) fanout.
The traces are made of 1 \textmu m thick aluminium film deposited on a SiO$_2$ passivation layer by means of magnetron sputtering. 
A prototype sensor is shown in Fig.~\ref{sensor_GAAS_pic}.
\begin{figure}[ht!]
\begin{center}
  \includegraphics[width=0.85\columnwidth]{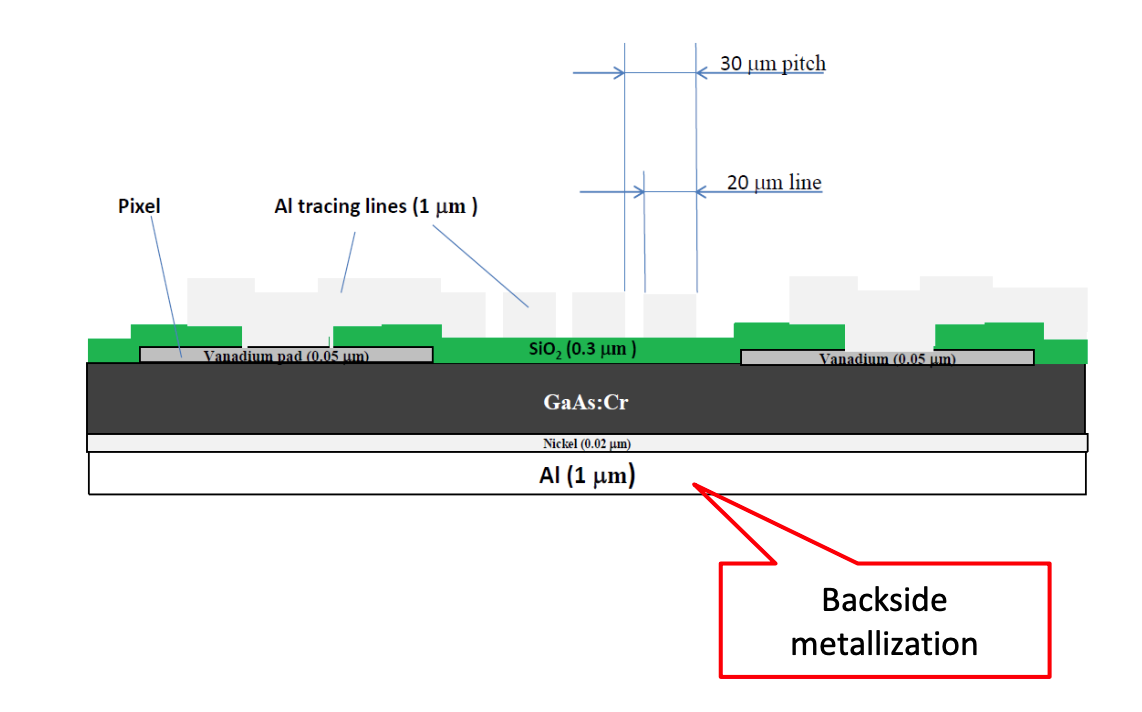}
    \caption{Schematic cross-profile of a GaAs sensor, not in scale. The aluminium traces are positioned between the columns of pads, on top of a SO$_2$ passivation layer.}
    \label{sensor_GAAS_cut}
\end{center}
\end{figure}
An illustration of the structure, as described above, can be seen in the cross-profile shown in Fig.~\ref{sensor_GAAS_cut}.
The leakage current of all pads was measured as a function of the bias voltage. A typical example is shown in Fig.~\ref{sensor_GaAs_leakage}. The leakage current rises almost linearly with the bias voltage, as expected for a compensated semiconductor. At a bias voltage of $100 \units{V}$ the leakage current amounts to about $50 \units{nA}$. 

A flexible Kapton PCB is used to connect the sensor pads to the front-end ASIC board (FEB). In the case of the Si sensor, copper traces on the PCB are connected to the pads with the conductive glue Epotek 4110. 
 For the GaAs sensor, the bonding pads on the top of the sensor, as shown in Fig.~\ref{sensor_GAAS_pic}, are connected to the PCB.
\begin{figure}[ht!]
\begin{center}
\includegraphics[width=0.99\columnwidth]{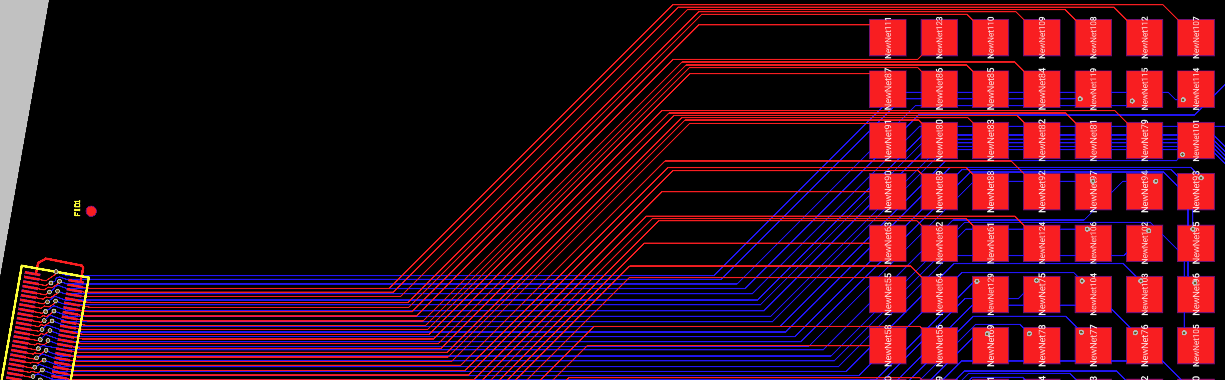}
\caption{Design of the flexible PCB used to read out the Si sensor. Bottom left, one sees the location of the connector. The red squares will be glued to the sensor pads. The red and blue lines represent the traces between pads and the connector pins (in red, traces located on the top layer and in blue traces located on the bottom layer of the flexible PCB).}
\label{pcb}
\end{center}
\end{figure}
A picture of the PCB used for the Si sensor read out is shown in 
Fig.~\ref{pcb}.
For both sensors, the bias voltage is supplied to the sensor through  
Kapton PCB glued to the sensor back-plane.

\section{Front-end Electronics and Data Acquisition}\label{sec3}

\begin{figure}[htbp]
 \begin{center}
  \includegraphics[width=0.93\textwidth]{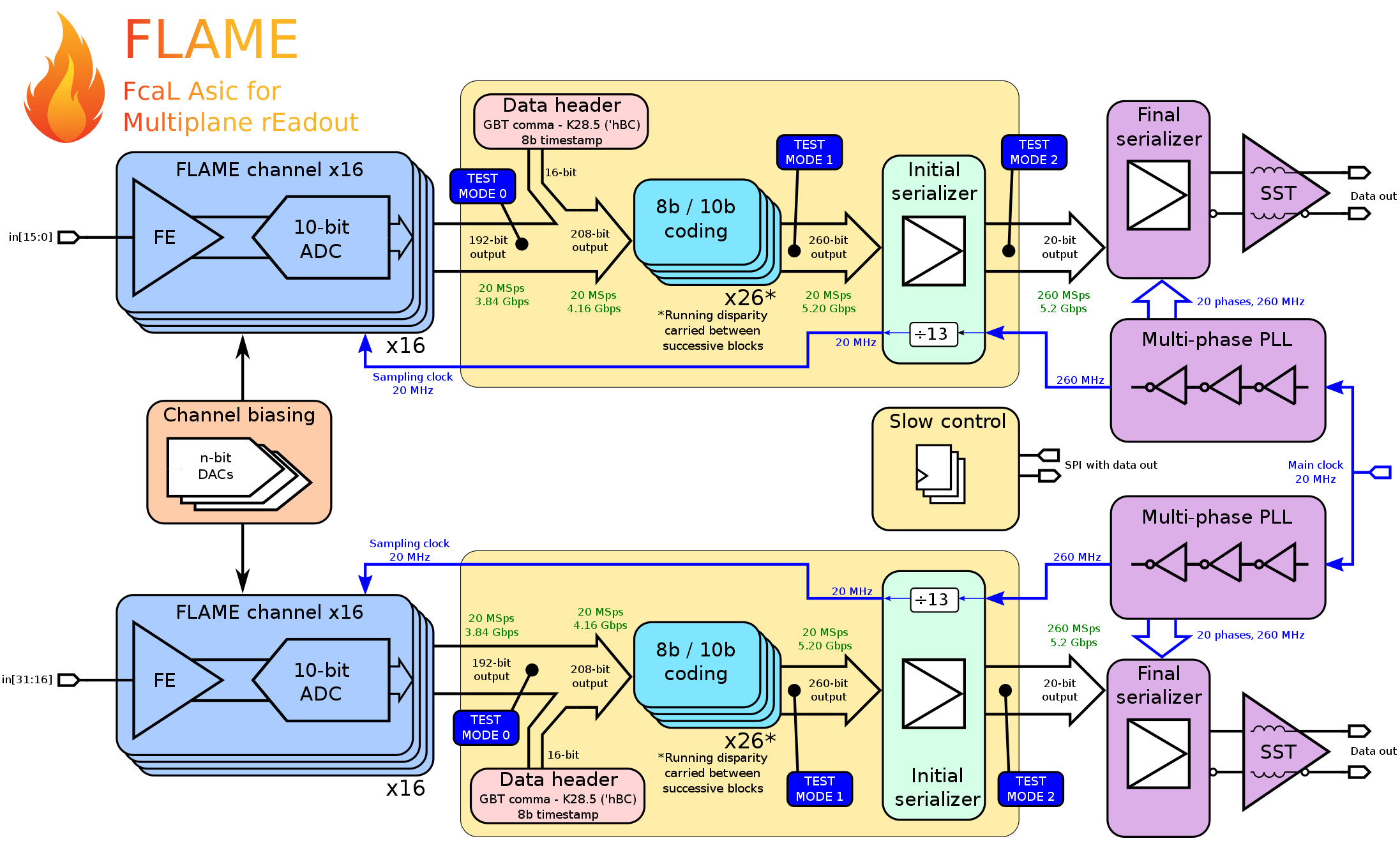}
    \caption{Block diagram of a 32-channel FLAME ASIC.} 
    \label{fig:FE_flame}
    \end{center}
\end{figure}

Each sensor plane is read out by front-end (FE) ASICs called FLAME (\textbf{F}ca\textbf{L} \textbf{A}sic for \textbf{M}ultiplane r\textbf{E}adout), designed for silicon-pad detectors of the LumiCal calorimeter for a future electron-positron linear collider experiment~\cite{FLAME_1, FLAME_2}.
The main specifications of the FLAME ASIC are shown in Table~\ref{tab:flame}.
\begin{table}[htbp]
\begin{center}
\begin{tabular}{|l|l|}
\hline
Variable & Specification \\
\hline
Technology & TSMC CMOS 130~nm \\
Channels per ASIC & 32 \\
Power dissipation/channel & $ 3.1 \units{mW}$ \\
\hline
Noise & $\sim$1000 e$^-$@10\,pF + 50e$^-$/pF \\
Dynamic range & Input charge up to $\sim 6 \units{pC}$ \\
Linearity & Within 5\% over the dynamic range\\
Pulse shape  &  T$_\text{peak} \sim 55~\mathrm{ns}$\\
\hline
ADC bits & 10 bits  \\
ADC sampling rate & up to $\sim 20 \units{MSps}$ \\
Calibration modes & Analogue test pulses, digital data loading\\
Output serialiser & serial Gb-link, up to 9\,Gbit/s\\
Slow controls interface & I\textsuperscript{2}C, interface single-ended\\
\hline
 \end{tabular}
 \end{center}
 \vspace*{0.3cm}
 \caption{Summary of the specifications of the FLAME ASIC.}
 \label{tab:flame}
\end{table}
A block diagram of FLAME, a 32-channel ASIC designed in CMOS 130\,nm technology, is shown in Fig.~\ref{fig:FE_flame}. FLAME comprises an analogue FE and a 10-bit ADC in each channel, followed by a fast data serialiser. It extracts, filters and digitises analogue signals from the sensor, performs fast serialisation and transmits serial output data.
As seen in Fig.~\ref{fig:FE_flame}, the 32-channel chip is designed as a pair of two identical 16-channel blocks. Each block has its own serialiser and data transmitter so that during operation, two fast data streams are continuously sent to an external data acquisition system (DAQ). The biasing circuitry is common to the two 16-channel blocks and is placed in between. Also the slow control block is common and there is only one on the chip.
The analogue FE consists of a variable gain preamplifier with pole-zero cancellation (PZC) and a fully differential CR--RC shaper with peaking time of about $55 \units{ns}$. 
The shaper includes also an 8-bit Digital to Analog Converter (DAC), with $32 \units{mV}$ range, for precise baseline setting. The analogue FE consumes in total 1.5\,mW/channel.
The Analogue to Digital Converter (ADC) digitises with 10-bit resolution and at least 20\,MSps sampling rate. 
Using asynchronous readout, the sampling rate should be high enough to reconstruct the signal amplitude from the pulse samples. In standard readout mode, fast deconvolution is performed in the FPGA using a procedure that,
based on the first-order semi-Gaussian pulse shape, requires three samples~\cite{Kulis:2011zz}. The same procedure is applied to raw data, and sometimes a
standard fit is also used for verification. The reconstruction quality
works well when the sampling period is similar to the peaking time of the semi-Gaussian pulse.
The power consumption is below 0.5\,mW per channel at 20\,MSps. 
In order to ensure the linearity of the ADC, the input switches are bootstrapped, significantly reducing their dynamic resistance.
The dynamic range of the ASICs can be switched between high and low gain. At high gain, the response to the input charge is almost linear between
0.5 to 75  
minimum-ionising-particle (MIP) equivalent depositions in a 500\,\textmu m thick Si sensor. At low gain, up to a charge of $5\units{pC}$ the response is linear, corresponding to about 1000 MIP-equivalents\footnote{For input charges between $5\units{pC}$ and $50\units{pC}$ the response is non-linear, hence a dedicated calibration is needed to determine the size of the input charge.}, and the lower threshold is in the range of a few 10 MIP-equivalents.
Currently, the gain setting cannot be changed
dynamically, i.e. the existing implementation would require the gain setting to be changed depending on the position of the readout ASIC in the electromagnetic shower. A more flexible implementation with dynamic gain change is
foreseen in the future.

Data from the ASICs are collected and pre-processed in a back-end FPGA board. 
When a trigger
signal is sent to the FPGA, up to 64 raw ADC samples are collected in an event for each of the readout channels.
In the raw data readout mode, all ADC samples are recorded. In the standard readout mode, the event is processed by the FPGA as described above. For signals on pads above a predefined threshold, the signal size and the time-of-arrival (TOA) are calculated and recorded. This procedure, combined with zero suppression, significantly reduces the amount of data. Finally, the event data are sent from the FPGA board to the DAQ computer using the User Datagram Protocol (UDP) through a single 1\,Gbps Ethernet link. 

\begin{figure}[htb!]
    \centering
    \includegraphics[width=\textwidth]{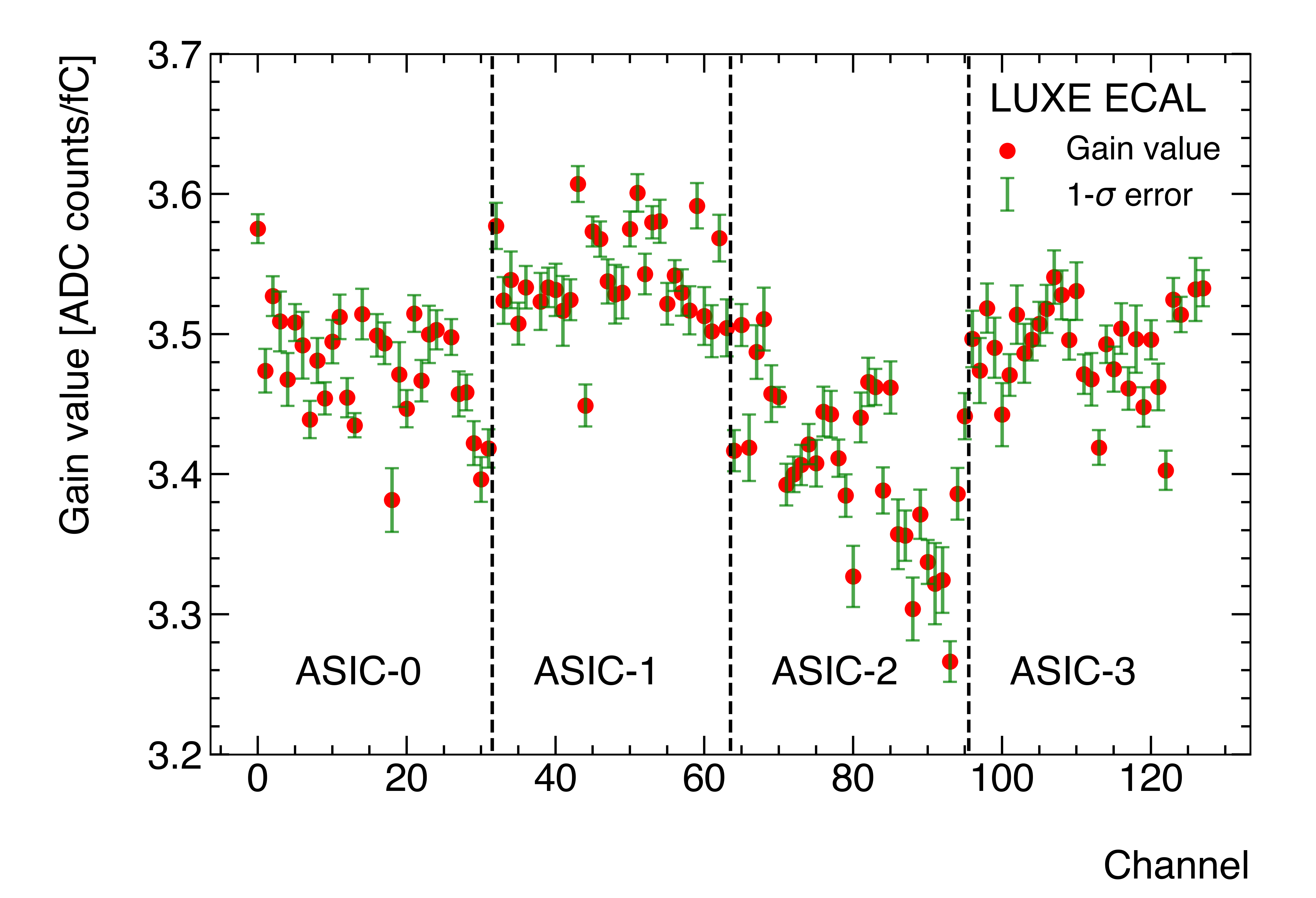}
    \caption{Measured gain of all channels using a charge injector. The dashed lines indicate the ranges of different ASICs.}
    \label{fig:gain_ch}
  \hspace{0.01\textwidth}
   \centering
    \includegraphics[width=\textwidth]{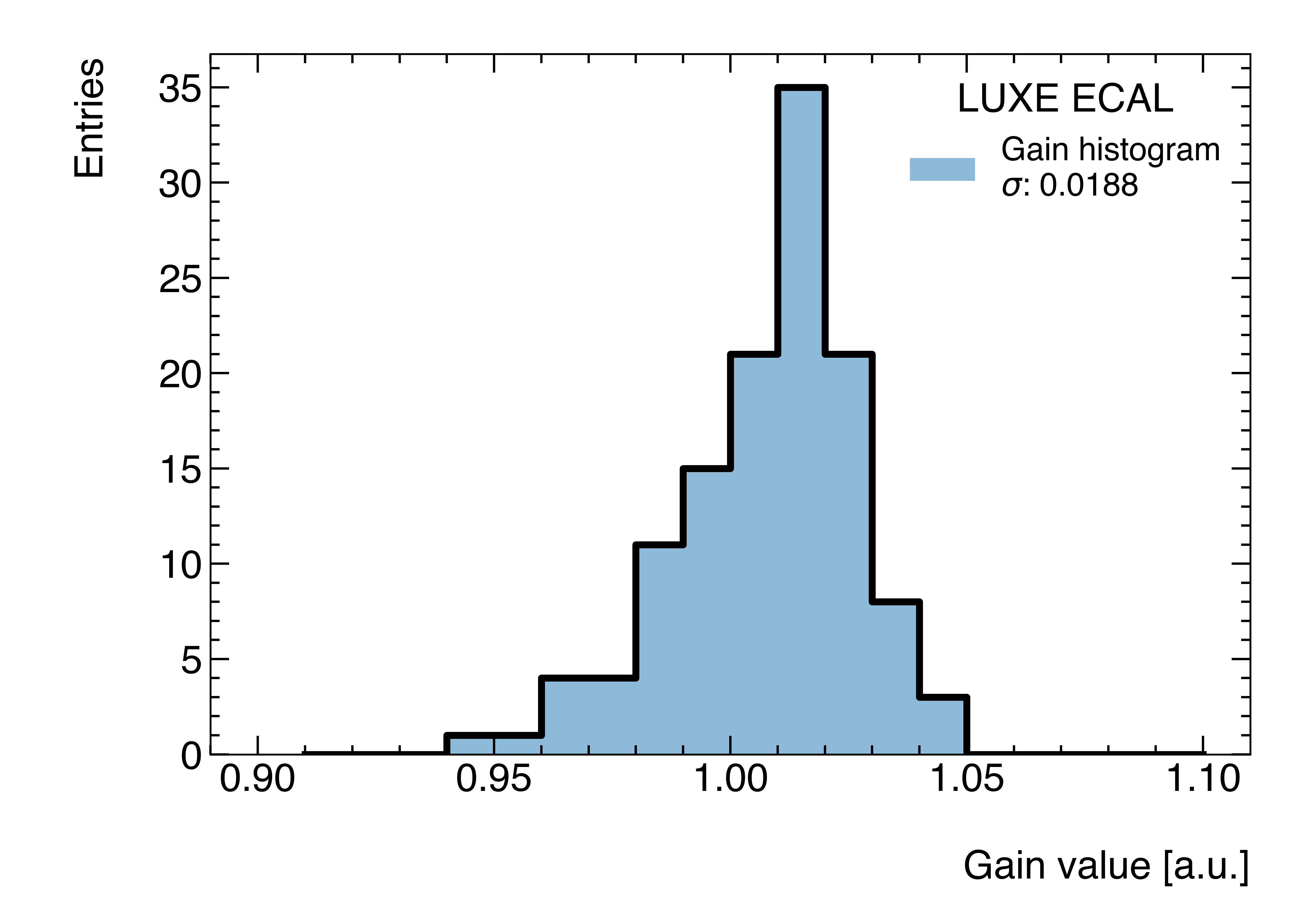}
    \caption{The distribution of the normalised gains. }
    \label{fig:gain_hist}
\end{figure}

To correct for differences in the amplification of the FE amplifiers, each readout channel is calibrated using a capacitor-resistor network to feed-in a known test-charge in the preamplifier input. The result for all channels in terms of ADC counts per fC is shown in Fig.~\ref{fig:gain_ch}. The distribution of the gains, normalised to the mean value, for all readout channels is shown in Fig.~\ref{fig:gain_hist}.
As can be seen, the normalised gains vary within 1.9\%.
 
\section{Beam, Trigger and Beam Telescope }\label{sec4}

\begin{figure}[h!tbp]
 \begin{center}
  \includegraphics[width=\textwidth]{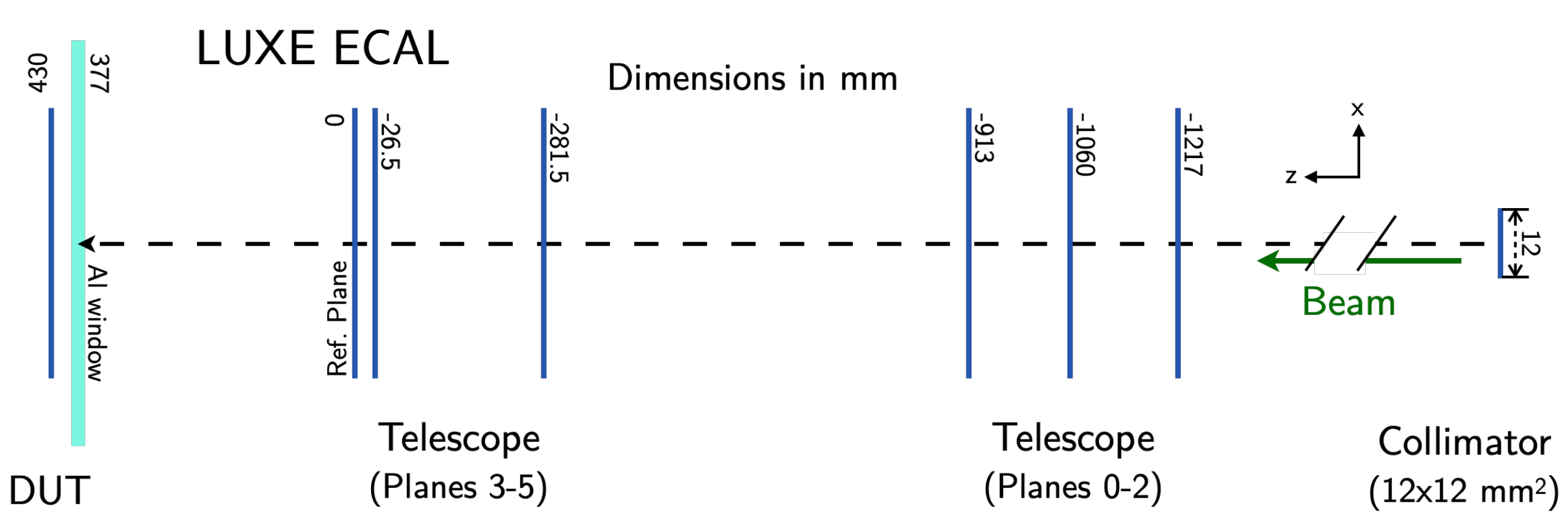}
    \caption{Sketch of the test beam set-up. Electrons arrive from the right, pass the first scintillator, then six Alpide sensor planes, the second scintillator, and hit the sensor, denoted here as DUT. The dimensions are given in mm.} 
    \label{fig:TB_set-up}
    \end{center}
\end{figure}

\begin{figure}[h!tbp]
 \begin{center}
  \includegraphics[width=0.9\textwidth]{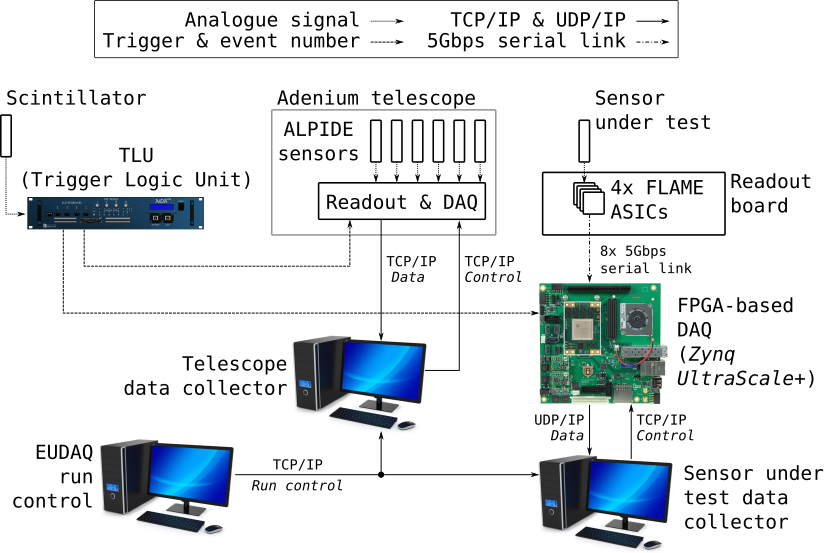}
    \caption{Scheme of the readout system. The TLU trigger is sent both to the telescope and to the FPGAs. The FPGAs orchestrate the FE ASICS and perform the pre-processing of the ADC raw data. The telescope and the FPGAs are read out by separate computers, monitored via the EUDAQ run control.} 
    \label{fig:read_out}
    \end{center}
\end{figure}

Electrons of 5\,\gev\ energy produced at the DESY-II test-beam facility~\cite{test-beam} are used in this study. 
The electrons pass a collimator of $1.2 \times 1.2 \units{cm^2}$ aperture. Two scintillation counters upstream and downstream of the beam telescope are used to form a trigger signal in the Trigger Logic Unit (TLU)~\cite{Baesso:2019smg}. 
The beam telescope comprises six planes of Alpide sensors with a sensitive area of $1.5 \times 3.0 \units{cm^2}$ and a
pixel pitch of $29.24 \times 26.88$\,\textmu m$^2$. 
The sensors under test are positioned just downstream of the last telescope-plane. A sketch of the test beam set-up is shown in Fig.~\ref{fig:TB_set-up}.
Both the telescope sensors and the sensor under test are read out separately after arrival of a TLU trigger signal and form an event. The TLU delivers a trigger number for synchronising the records from the telescope and the sensor. In addition, a time stamp is given to each record. The readout flow is monitored by the EUDAQ~\cite{ Ahlburg:2019jyj,Liu:2019wim} run control.
The average trigger rate during data taking was about 1.5\,kHz. The scheme of the readout is shown in Fig.~\ref{fig:read_out}.

\begin{figure}[htbp]
 \begin{center}  
 \includegraphics[width=0.7\textwidth]{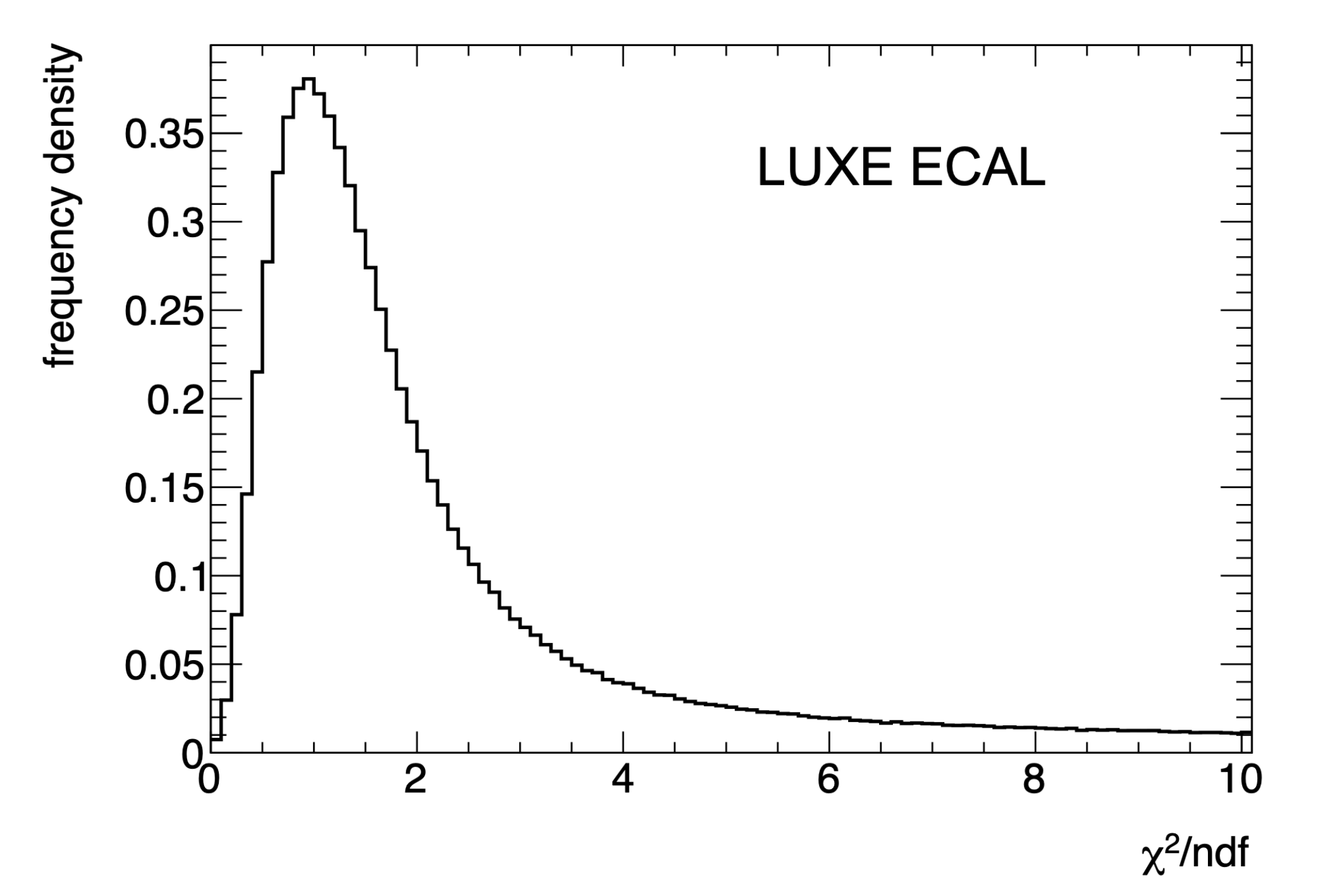}
    \caption{ The distribution of $\chi^2/\text{ndf}$ ($\chi^2$ per number of degrees of freedom) for electron trajectories fitted in the telescope. } 
    \label{fig:TB_chi2}
    \end{center}
\end{figure}

The beam telescope is used to measure the trajectory of each beam electron. The track reconstruction is done with the software package Corryvreckan using the General Broken Line option~\cite{Dannheim:2020jlk}. For the alignment of the telescope planes, about 50 k events at the beginning of each run are used. A typical $\chi^2$ distribution of the track fit is shown in Fig.~\ref{fig:TB_chi2}.
The uncertainty of the prediction of the impact point of the electron on the sensor plane amounts to about 37\,\textmu m, dominated by multiple scattering in the downstream telescope plane and the air-gap between this plane and the sensor.

\section{Data Taking and Analysis}\label{sec5}

\subsection{Detector alignment}\label{subsec1}
In order to perform a systematic scan of the signal response of sensor pads for each run,
the location of the sensor pads in the coordinate system of the telescope has to be determined from data. 

\begin{figure}[ht]
    \centering
    \includegraphics[width=0.495\textwidth]{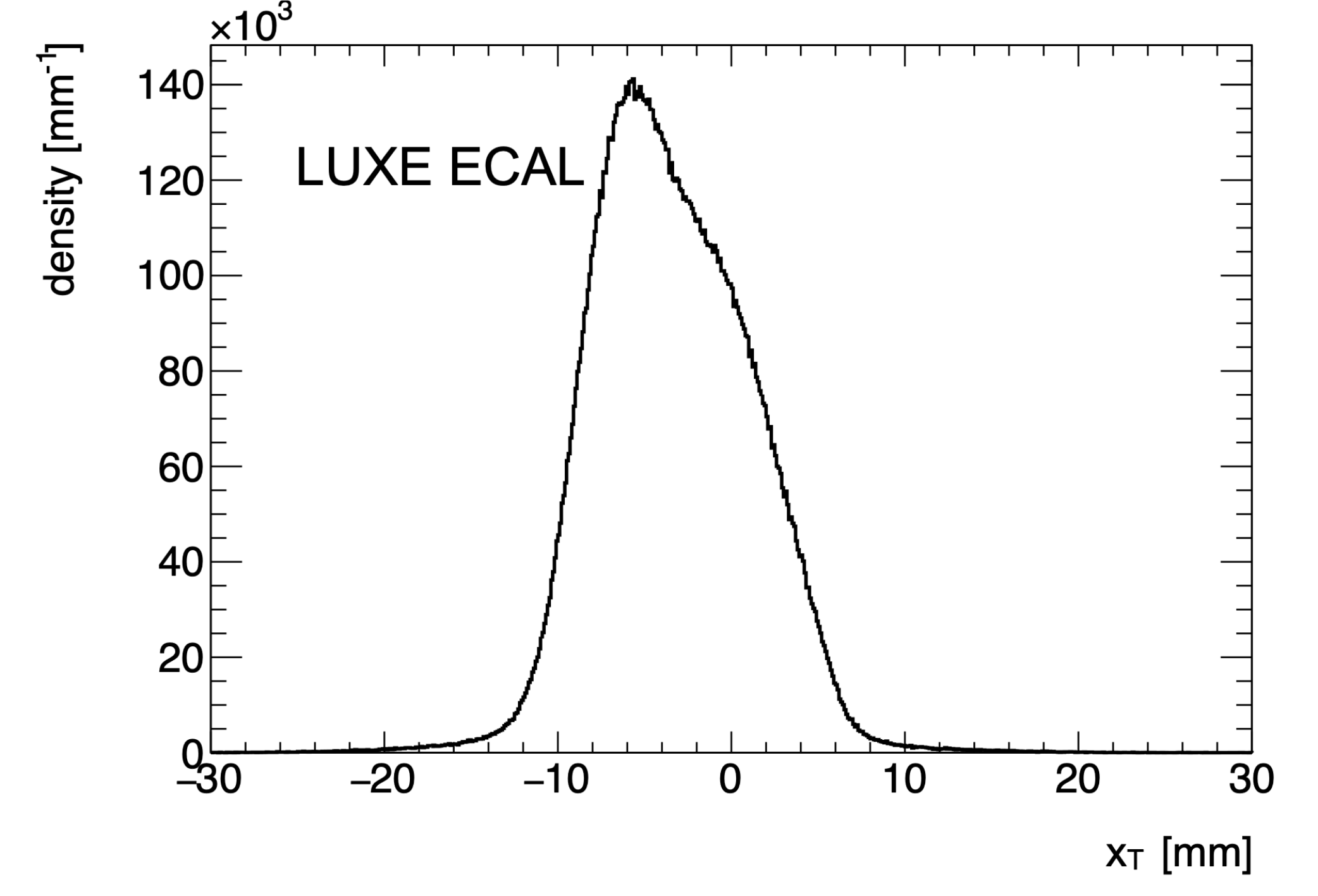}
    \includegraphics[width=0.495\textwidth]{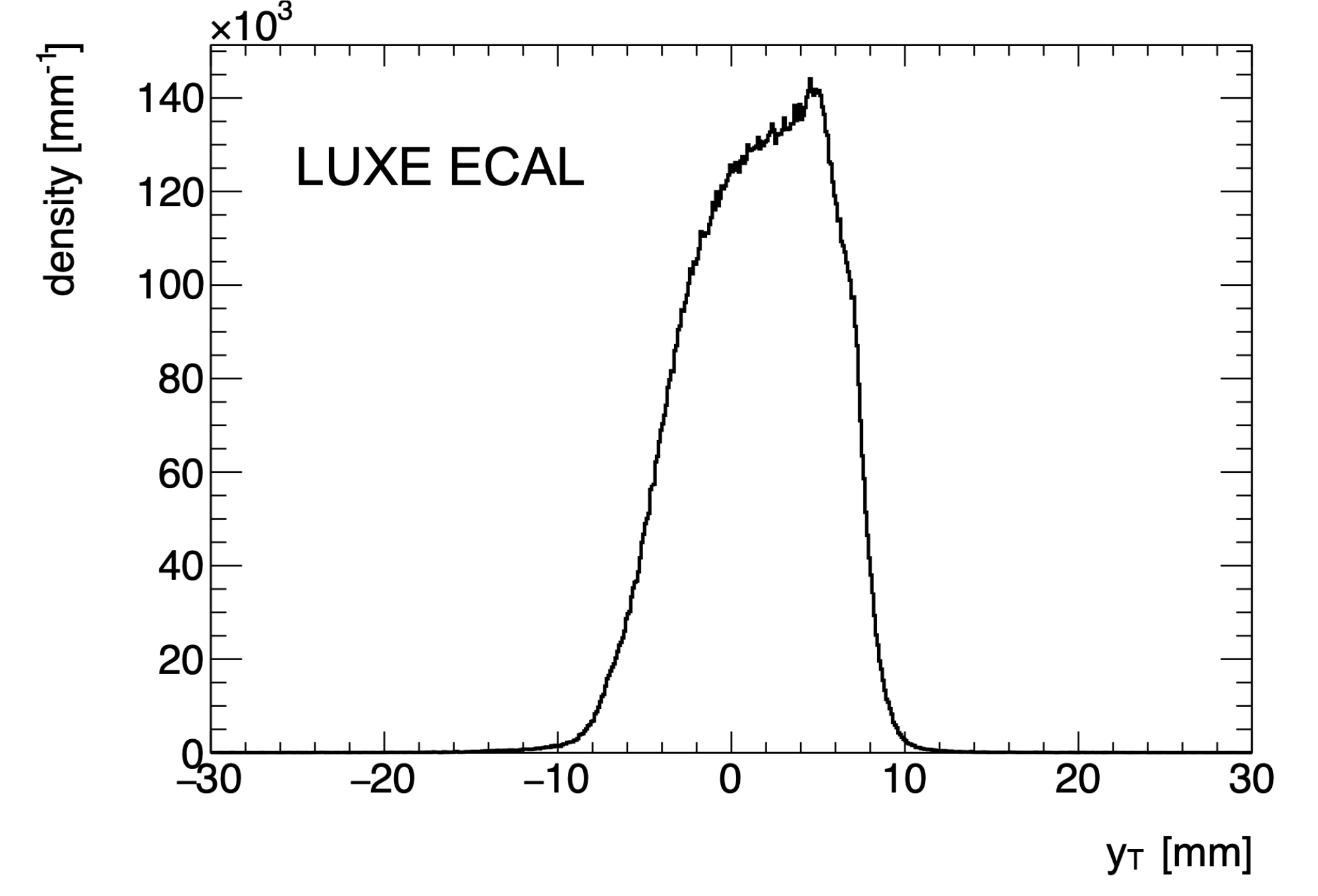}
    \caption{Projection of the beam profile on the $x$-axis (left) and $y$-axis (right) of the sensor plane using telescope data.}
    \label{fig:beam-profile}
\end{figure}

\begin{figure}[ht]
    \includegraphics[width=0.495\textwidth]{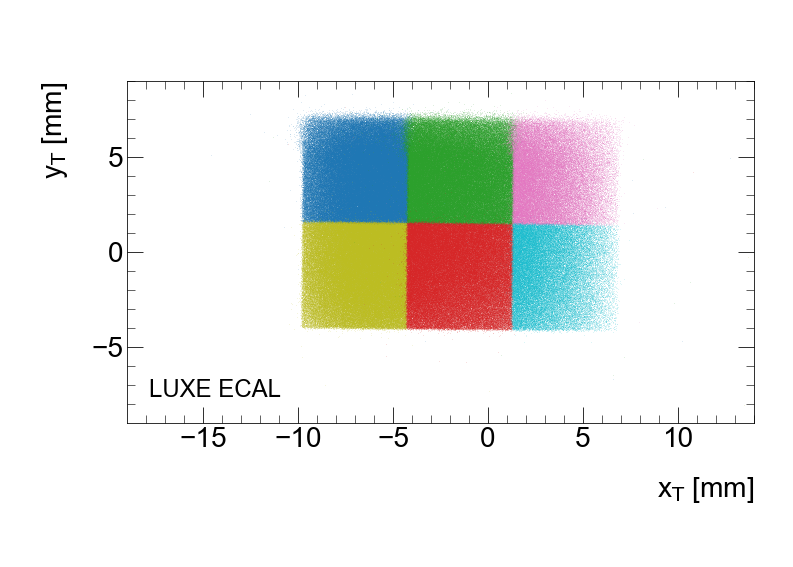}
    \includegraphics[width=0.495\textwidth]{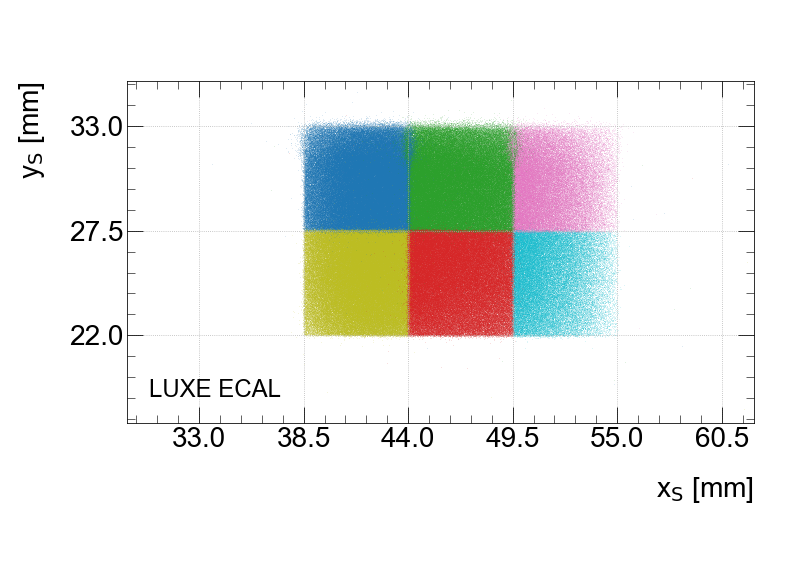}
    \caption{
      The example maps of expected track position on the face of the Si sensor in the telescope coordinate system, $x_\mathrm{T}$ and $y_\mathrm{T}$ before alignment (left) and
       in the sensor coordinate system, x$_\mathrm{S}$ and y$_\mathrm{S}$, after alignment (right).
        Each pad is assigned the same colour in both maps.
    }
    \label{fig:IL-hits-before-alignment}
\end{figure}

The profile of the beam at the sensor plane, shown in Fig.~\ref{fig:beam-profile}, is such that at maximum an array of $3 \times 3$ pads is exposed to the beam.
Two complementary procedures for the telescope-sensor alignment were developed.
In both cases, only the translation and rotation of the $x$ and $y$ sensor coordinates in the telescope system is considered. 
The distance between the telescope reference plane and the sensor plane was measured in situ.
The issue whether the two planes are parallel can be addressed after the translation and rotation is performed, by checking whether the parameters of the sensor-grid conform to the expectations.
The first alignment procedure consists of finding the transformation for which the number of hits observed in a given pad is maximised relative to the expected position of the hits from the telescope.
For this procedure, a sample of events is selected with only a single pad with signal compatible with a minimum ionising particle as well as a single track reconstructed in the telescope. Each pad is then assigned a ``colour'' and the map of the expected $x$ and $y$ positions extrapolated from the reconstructed track is drawn in that colour.
An example is shown in Fig.~\ref{fig:IL-hits-before-alignment}.
A colour grid compatible with the pad structure of the sensor is observed. A slight rotation of the sensor in the telescope coordinate system is detected. There are also hits of a given colour which are predicted outside the pad they were assigned to. These hits indicate that the extrapolation of the track to the sensor is affected by the telescope resolution and, in addition, by multiple scattering in the downstream telescope plane and the air-gap between this  plane and the sensor. 
The number of hits observed outside the assigned pad can be used to estimate the precision with which the expected position of the hit is determined. A model assuming a Gaussian smearing and the measured beam profile leads to a resolution of 40\,\textmu m. This number, obtained after eliminating large values of $\chi^2/\text{ndf}$ to remain compatible with a Gaussian behaviour, is in agreement with the estimate obtained  from the telescope inherent resolution combined with the impact of multiple scattering. 

The second procedure makes use of the edge effects between sensor pads. The idea is to use beam-electrons entering the inter-pad area and to apply a Hough transform~\cite{Duda:1972ymn} to find the lines that correspond to the pad edges. A map of the grid-edges in the telescope coordinate system is obtained  by selecting events with a single track, and exactly two adjacent pads with signals, either in the horizontal or vertical direction. 
\begin{figure}[ht]
  \begin{center}
    \includegraphics[width=0.495\textwidth]{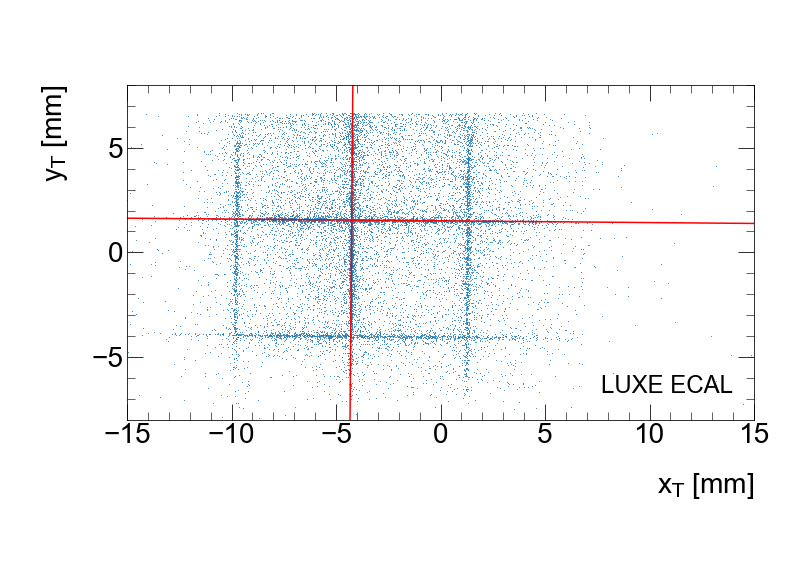}
  \caption{The map of expected track position on the face of the Si sensor in the telescope coordinate system, $x_\mathrm{T}$ and $y_\mathrm{T}$, for events in which the beam is expected to have crossed the inter-pad area. Also shown are the lines obtained by applying the modified Hough transform.}
  \label{fig:ME-edges grid}
 \end{center}
\end{figure}
A modified Hough transform~\cite{1990PaReL11167B} is used to determine the position of the grid-lines. 
The parameter space consists of the vertical shift and rotation of the sensor coordinate system in the telescope system.
To minimise the combinatorial background in the line parameter-space, the parameters are determined for each pair of hits, provided the distance between the two hits is sufficiently large in $x$ ($y$) for horizontal (vertical) grid-lines. The parameters of the lines will show up in the parameter space as maxima. The results of this procedure are shown in Fig.~\ref{fig:ME-edges grid}. The distribution around these maxima can be used to determine the precision of the alignment.

The results obtained with the Hough transform are nearly identical to the ones from maximising the pad content, however this procedure is faster and therefore easier to apply for alignment in the analysis. 

\section{Results}\label{sec6}

\subsection{Signal Size Distributions}\label{subsec2}

\begin{figure}[ht!]
    \includegraphics[width=0.48\textwidth]{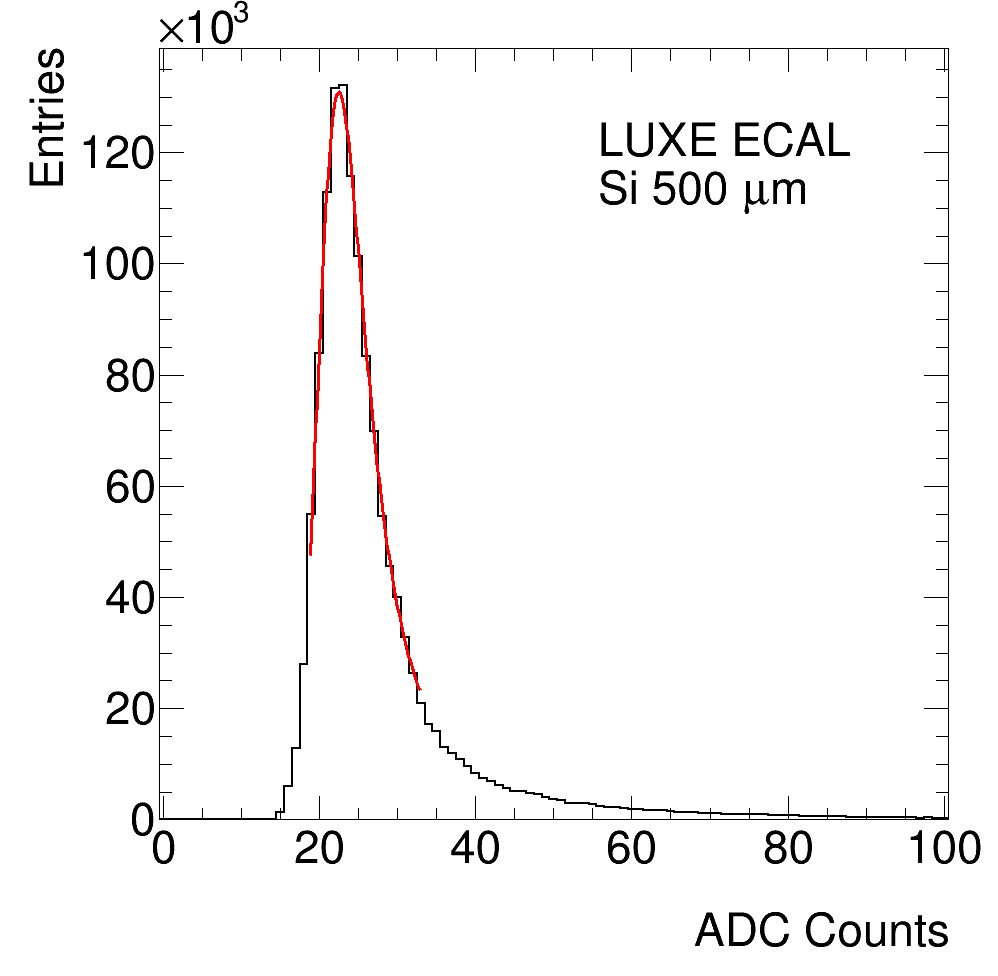}
    \includegraphics[width=0.48\textwidth]{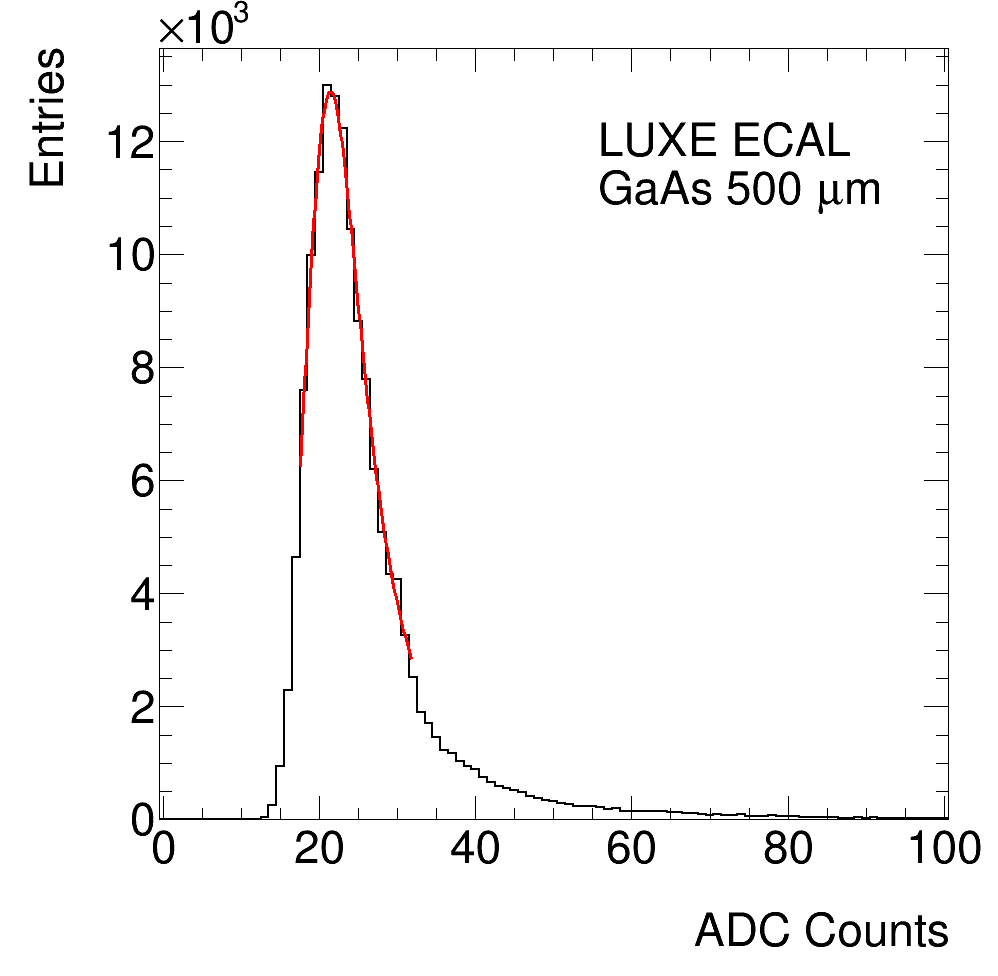}
    \caption{Distribution of the signal in a pad of the Si (left) and GaAs (right) sensor. The continuous line represents the result of a fit with a Landau distribution function convoluted with a Gaussian in the corresponding range. }
    \label{fig:signal_silicon_GaAs}
\end{figure}

Using tracks of the telescope with a predicted impact point on the area of a pad with the small regions near the edges excluded, the signal recorded from the DAQ in units of ADC counts is obtained. Typical signal distributions are shown in Fig.~\ref{fig:signal_silicon_GaAs} 
for the Si and GaAs sensors, respectively.

 A fit is performed with a Landau distribution function convoluted with a Gaussian to determine the most probable value, MPV, resulting in values of 21.9 and 20.3 ADC counts for the Si and GaAs sensors, respectively. As explained below (see~\ref{subsec3}), the similarity between the MPVs for silicon and GaAs is incidental. Using the measured widths of the pedestals of 1.4 and 1.2 ADC channels, the signal-to-noise ratio is 15 and 17 for the Si and GaAs sensors, respectively. Hence, this system of sensor planes and readout is well suited to detect MIPs with high efficiency, being potentially important for alignment and channel-by-channel calibration. 

\subsection{Monte Carlo simulation}\label{subsec3}

The response of both Si and GaAs sensors to $5\units{GeV}$ electrons is obtained using a stand-alone application based on the Geant4 package~\cite{AGOSTINELLI2003250} for simulation and the ROOT framework~\cite{BRUN199781} for outcome evaluation.
The Geant4 implementation of the sensor geometry includes, for the Si sensor, the Kapton PCB board used to supply the bias voltage and the Kapton-copper fanout for the signal routing. The sensor is divided into $16\times 16$ pads with each pad marked as sensitive volume for the energy deposition collection. 
For the GaAs, 
along with the sensor geometry, the nickel and aluminium back-planes and the gaps between pads are implemented. 

To model the interaction of the beam electrons with the sensors' material, the standard $QGSP\_BERT$ physics list was used with the 'option 4' or $\_EMZ$ of electromagnetic physics and a range cut-off of 1\,\textmu m.
The beam parameters are set using GPS commands from \textsc{Geant4}, and the source was simulated mimicking the beam after the collimator, with a square shape of $1.2\times 1.2 \units{cm^2}$ placed at $3.27\units{m}$ upstream of the sensors. Each beam electron crosses the trigger scintillators and the six telescope planes. 

\begin{figure}[htbp]
 \begin{center}
  \includegraphics[width=0.7\textwidth]{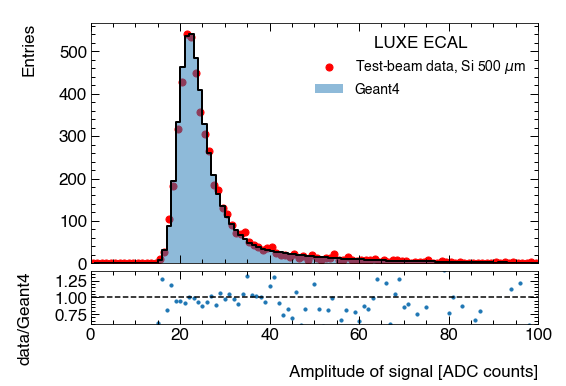}
    \caption{ Distribution of the Si sensor response to electrons, as measured in the test beam, compared to the Geant4 simulation results. The Monte Carlo distribution is normalised to the number of events in the maximum of the signal distribution obtained in data. The bottom panel shows the ratio of the two distributions.} 
    \label{fig:TB_MC1}
    \end{center}
\end{figure}

\begin{figure}[htbp]
 \begin{center}
  \includegraphics[width=0.7\textwidth]{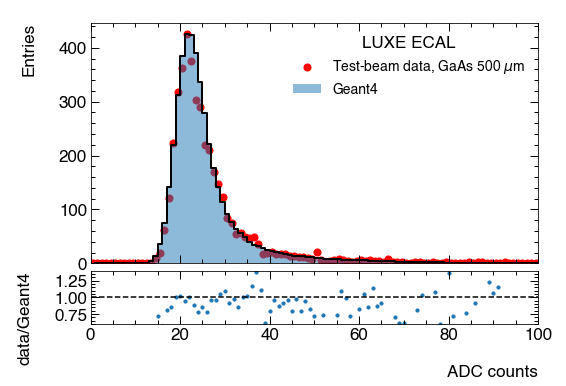}
    \caption{ Distribution of the GaAs sensor response to electrons, as measured in the test beam, compared to the Geant4 simulation results. The Monte Carlo distribution is normalised to the number of events in the maximum of the signal distribution obtained in data. The bottom panel shows the ratio of the two distributions.} 
    \label{fig:TB_MC2}
    \end{center}
\end{figure}

To model the response and the readout electronics, the following procedure is applied.
The energy loss of $5\units{GeV}$ electrons in the Si sensor is converted into the number of charge carriers using an average energy to create an electron-hole pair of $3.6$\,eV\ \cite{Fabjan:2020wnt}. 
The number of drifting charge carriers is then converted into ADC counts using the gain factor of the readout electronics, as measured in the laboratory with a known test-charge, of 3.47 ADC counts per fC. Gaussian smearing is applied to each readout channel to account for the noise of the FE electronics. The width of the Gaussian is obtained from a fit to pedestals measured in dedicated runs in the test-beam set-up. An additional correction factor is added as a free parameter in the fit in order to get agreement between the MPVs of the signal distribution in data and simulation. This factor amounts to 1.05.
The distribution of the signal size, as measured in the test-beam, is compared to the results of the \textsc{Geant4} simulation, after applying the readout model as described above, in Fig.~\ref{fig:TB_MC1}.
A very good modelling of the test-beam data is obtained.
The response of the GaAs sensor is simulated in a similar manner, and the comparison to data is shown in Fig.~\ref{fig:TB_MC2}. 
The 
energy to create an electron-hole pair is $4.2$\,eV\ \cite{PhysRevB.13.761}. The energy loss of electrons in GaAs is about a factor 2.1 larger than in Si. However, the product of hole mobility and hole lifetime is very small for chromium compensated sensors, such that the hole drift  contribution to the signal is highly suppressed \cite{AYZENSHTAT2002120} and assumed to be zero. The MPV values of the simulation and data agree again within 5\%. The nearly same values of the MPVs for Si and GaAs sensors of the same thickness are hence incidental.

\subsection{Response on Single Pads}\label{subsec4}

\begin{figure}[ht!]
    \includegraphics[width=0.45\textwidth]{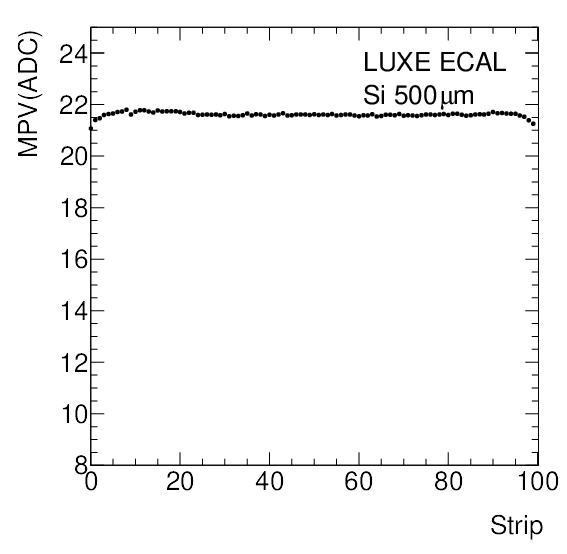}
    \includegraphics[width=0.45\textwidth]{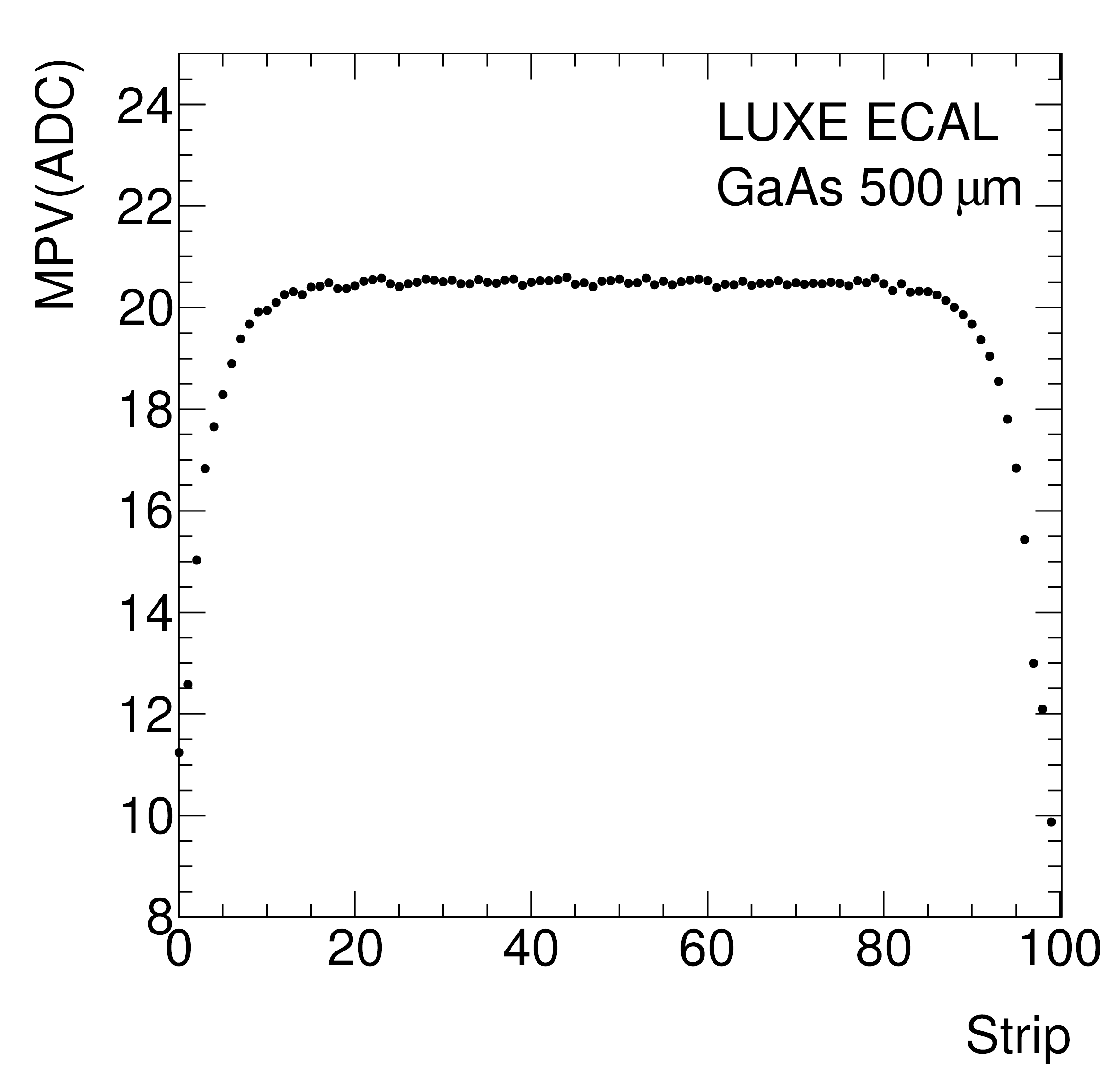}
    \caption{Distribution of the MPV of the signal of a Si sensor (left) and a GaAs sensor (right) as a function of the strip number along the $y$ coordinate.}
    \label{fig:signal_over_strip_silicon_GaAs}
\end{figure}

\begin{figure}[ht!]
    \includegraphics[width=0.45\textwidth]{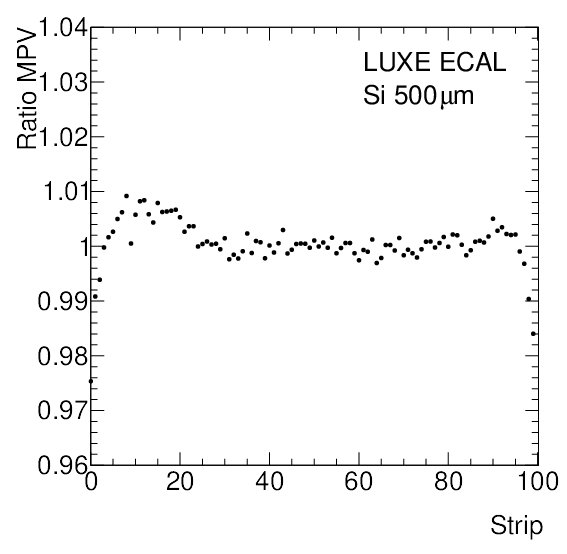}
    \includegraphics[width=0.45\textwidth]{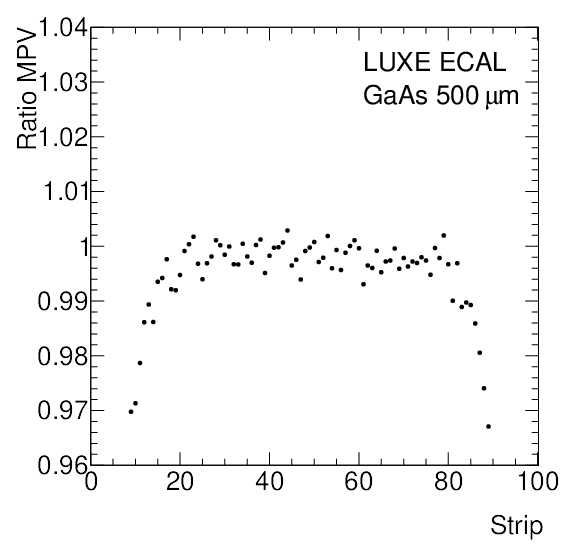}
    \caption{Distribution of the normalised signal size of a Si sensor (left) and a GaAs sensor (right) as a function of the strip number along the $y$ coordinate.}
   \label{fig:normsignal_over_strip_silicon_GaAs}
 \end{figure}

To study the uniformity of the response within a single pad, it is subdivided into virtual strips in $x$ and $y$ directions, with a strip width of 55 and 50\,\textmu m \footnote{These values are arbitrarily chosen but are in the range of the expected precision of the impact point predicted by the telescope.} for Si and GaAs sensors, respectively.
The MPV value is then determined for the signal distribution in each strip, each of which looks like previously shown in Fig.~\ref{fig:signal_silicon_GaAs}. 
A typical example of the MPV values obtained for the strips in the $y$ coordinate are shown as a function of the strip number in Fig.~\ref{fig:signal_over_strip_silicon_GaAs}, for both the Si and the GaAs sensors.
As can be seen, the signal size is stable over a large $y$ range while it drops down at the edges. For the GaAs sensor the signal drop is more pronounced than for the Si sensor, starting already about half a mm before the edge.
More details are visible when zooming into the ratio of the MPV to its central value within the pad. Examples are shown in Fig.~\ref{fig:normsignal_over_strip_silicon_GaAs}.
The MPV ratio as a function of the $y$-coordinate
for the Si sensor exhibits variations of about 1\%. For the GaAs sensor, a rapid signal drop $0.5 \units{mm}$ before the edges is confirmed. The same results are observed for the MPV dependence on the $x$ coordinate of a Si or GaAs sensor pad. 
These results may have impact on the precision of the shower position reconstruction and energy reconstruction, and should be carefully considered in performance simulations of a detector design.  

\subsection{Signal Size between Pads}\label{subsec5}

\begin{figure}[ht!]
    \centering
    \includegraphics[width=0.8\textwidth]{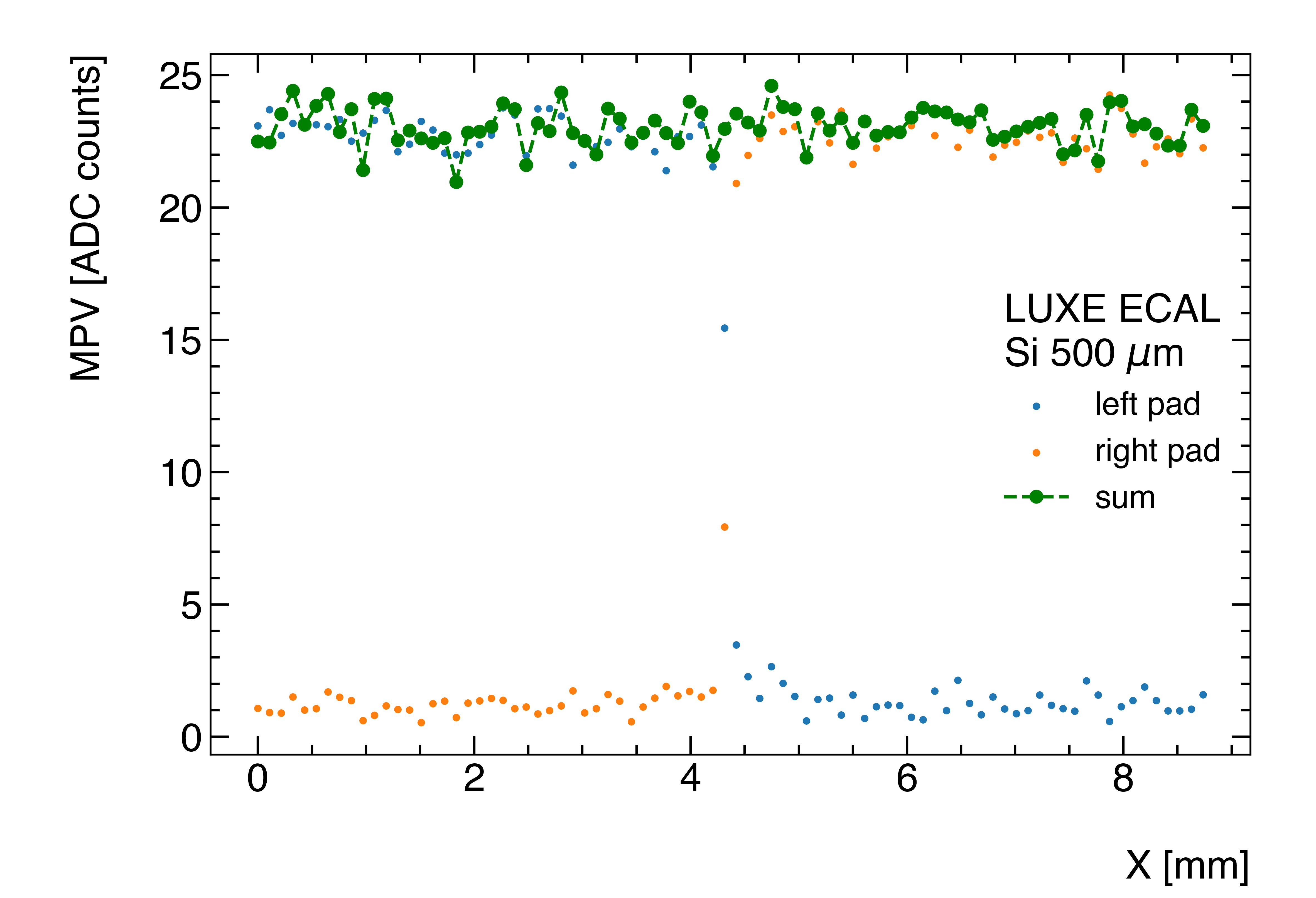}
    \caption{MPV as a function of the $x$ position in the region between neighbour pads for a Si sensor.}
    \label{fig:mean_4533_x}
\end{figure}

\begin{figure}[ht!]
   \centering
   \includegraphics[width=0.8\textwidth]{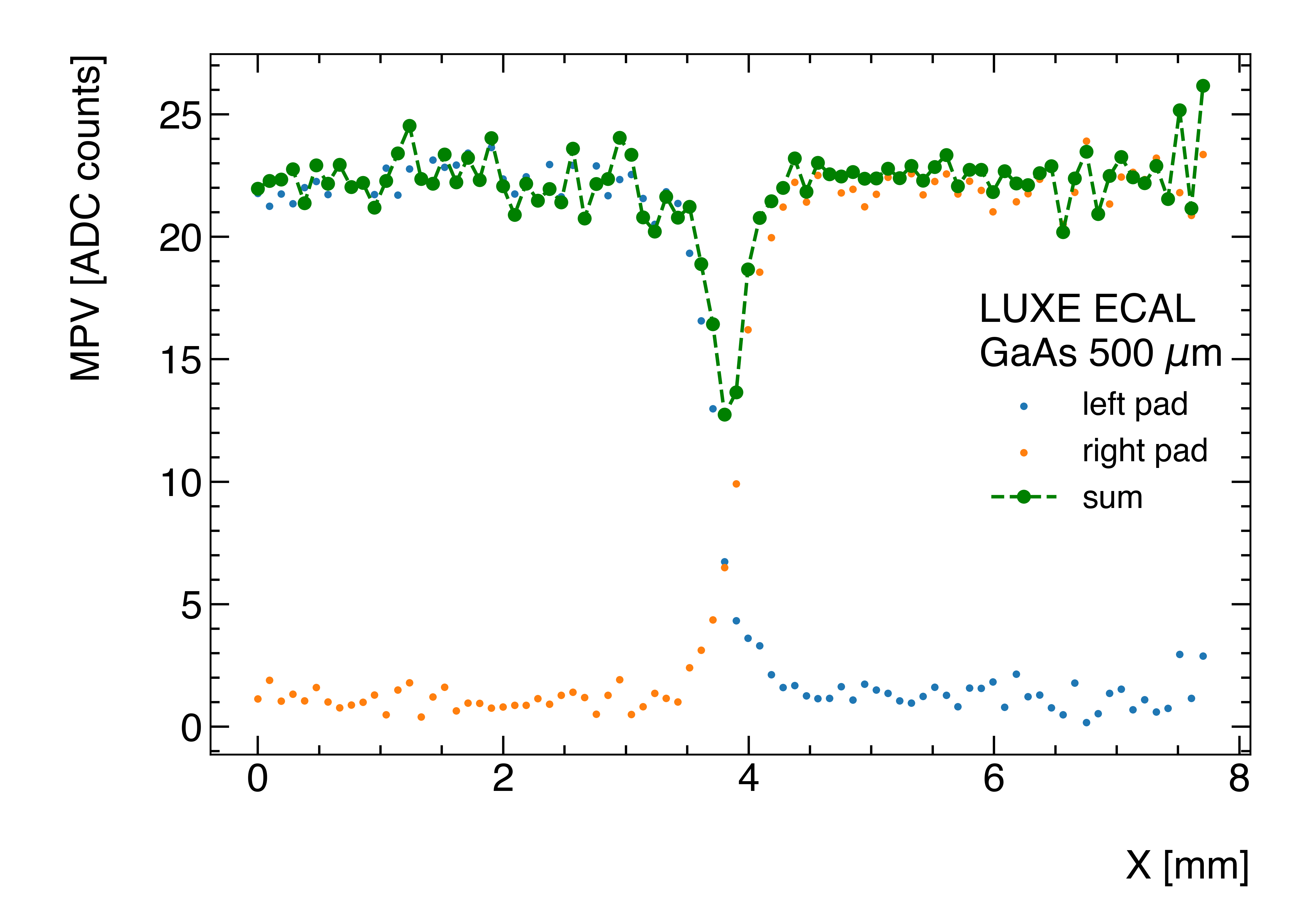}
    \caption{MPV as a function of the $x$ position in the region between neighbour pads for a GaAs sensor.}
    \label{fig:mean_4495_x}
\end{figure}

Using the impact point on the sensor as predicted by the telescope, the signal size was also studied in the regions between pads.
Examples for a Si and a GaAs sensor are shown in Figs.~\ref{fig:mean_4533_x}
and ~\ref{fig:mean_4495_x}, respectively.
The MPV is shown as a function of the local $x$  coordinate crossing the region between two pads. For the Si sensor, the MPV drops sharply at the edge of the pad while the MPV in the adjacent pad rises up. Adding at a given position both MPV values, no signal loss is observed in the transition from a pad to its neighbour.
The same result is obtained for the signal size as a function of the $y$ coordinate.
The same measurements for the GaAs sensor also show a drop of the response between adjacent pads. However, after adding the signal sizes of neighbour pads, measured at a certain position, the drop is still visible and amounts to about 40\%. For the GaAs sensor, the measurement as a function of the $y$-coordinate shows a drop of about 10\%. The drop is expected to be more pronounced in the $x$ coordinate due to the presence of the aluminium readout traces in the gap between pads.  
The relatively larger response drop in case of the larger gap may jeopardize the shower energy and position reconstruction in a calorimeter. A detailed simulation study should quantify the effect. A worse position resolution is particularly critical when defining the fiducial volume in a luminometer at electron-positron colliders~\cite{Abramowicz:2010bg}.

\subsection{Homogeneity of the Response  }\label{subsec6}

\begin{figure}[ht!]
\begin{minipage}[t]{0.5\textwidth}
    \centering
    \includegraphics[width=\textwidth]{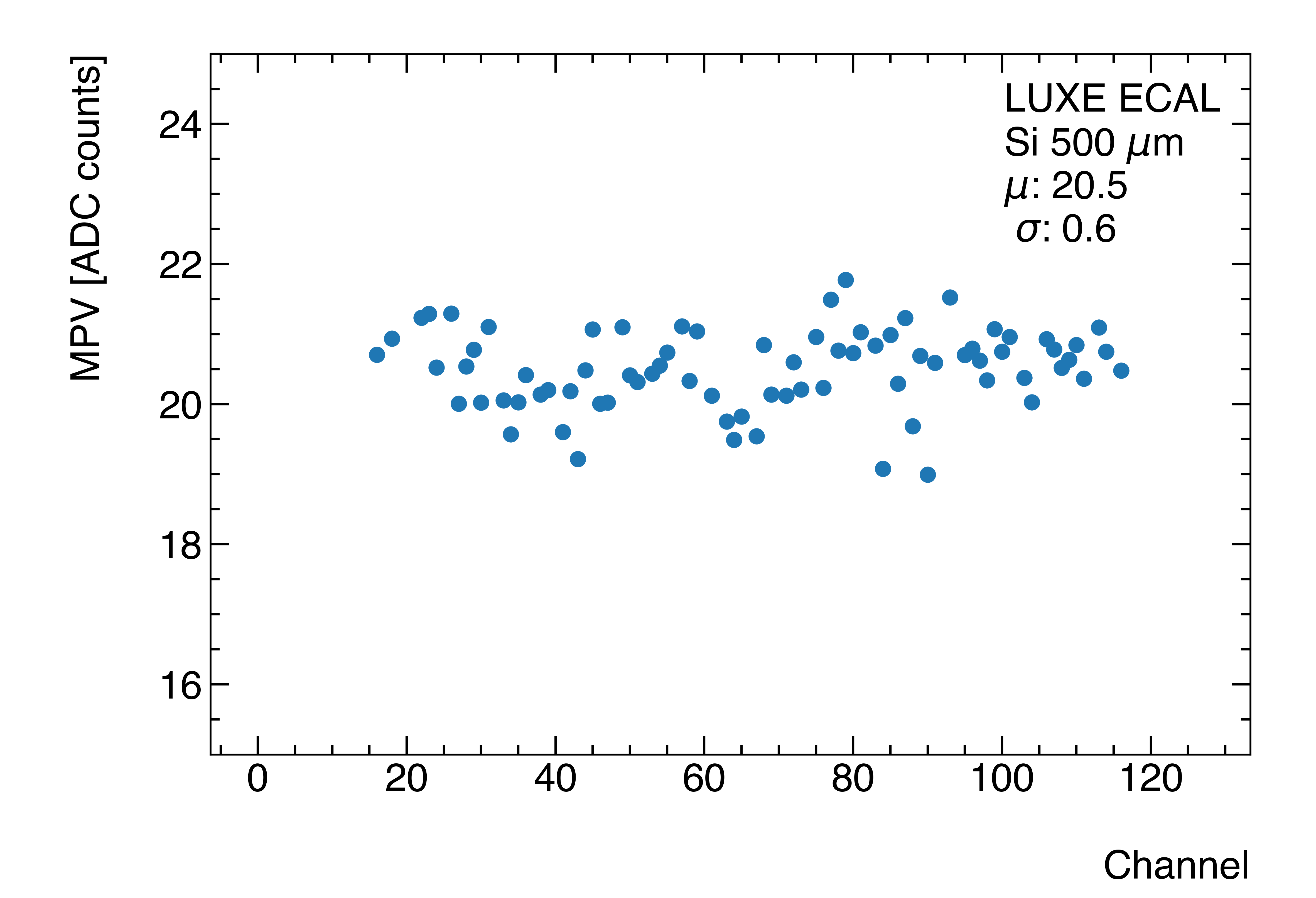}
    \caption{Distribution of the MPV values for all pads of a Si sensor as a function of pad (channel) number.}
    \label{fig:MPV_chanel}
  \end{minipage}\hfill
  \hspace{0.01\textwidth}
  \begin{minipage}[t]{0.5\textwidth}
   \centering
    \includegraphics[width=\textwidth]{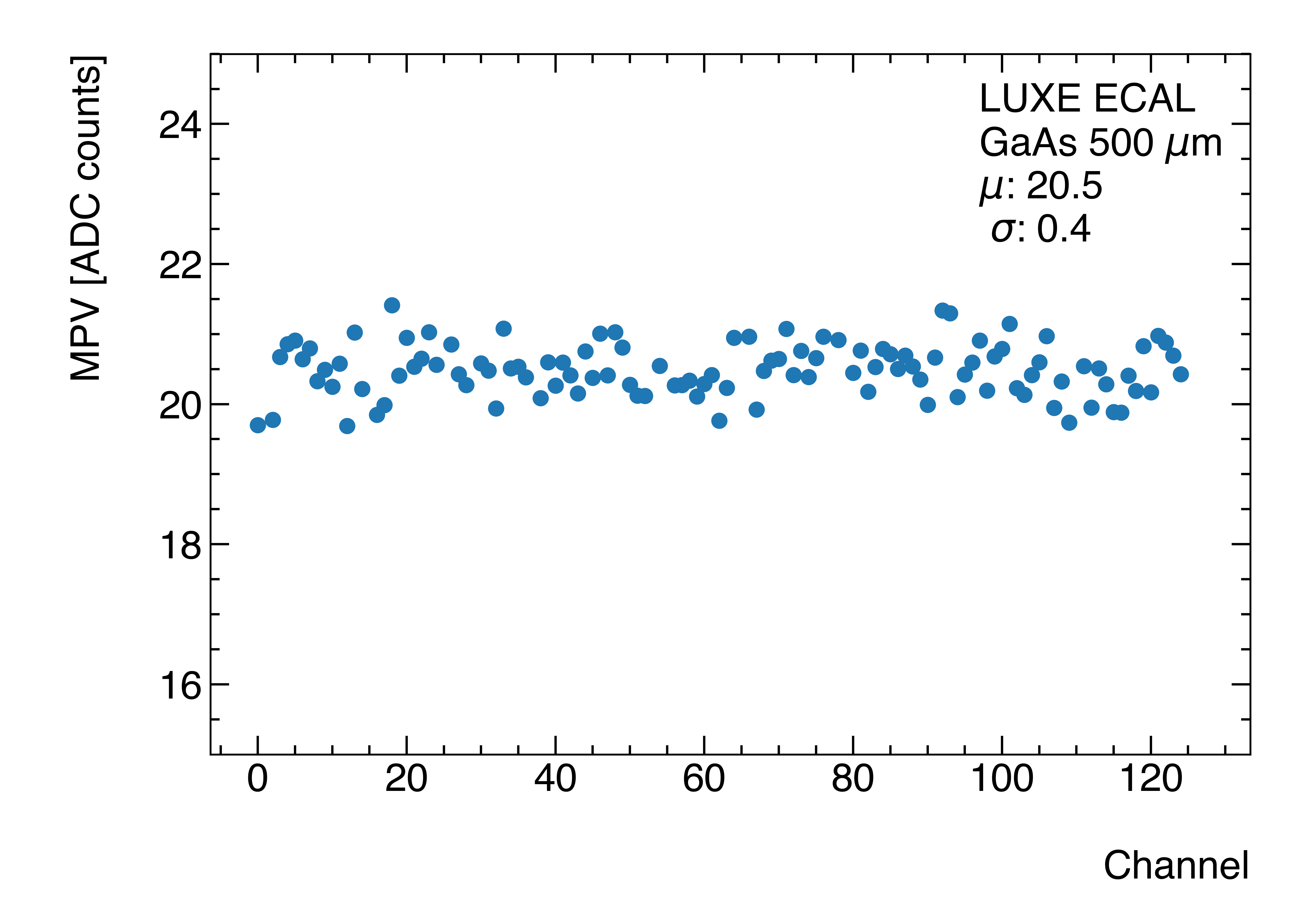}
    \caption{Distribution of the MPV values for all pads of a GaAs sensor as a function of pad (channel) number.}
    \label{fig:MPV_distribution}
  \end{minipage}\hfill
\end{figure}

By sweeping the beam spot over the whole surface of the sensors, the signal response of all pads was measured.
The values of the MPV determined for all pads of a Si and a GaAs sensor as a function of pad (channel) number are shown in Figs.~\ref{fig:MPV_chanel} and~\ref{fig:MPV_distribution}. The values are corrected for variations in the FE amplifier gains. 
The distribution of the MPV values for each sensor is almost Gaussian with mean values of 20.5 and 20.5 and widths of 0.6 and 0.4, in ADC counts, for the Si and GaAs sensor, respectively.
The measured spread of the pad response will be considered in ongoing simulation studies to understand whether it is critical for the performance of ECAL. In addition, the channel-by-channel gain differences are easily corrected for using test-beam data.

\subsection{Cross-talk studies }\label{subsec7}

For both Si and GaAs sensors, the capacitance between copper or aluminium strips is estimated to be less than $14 \units{pF}$. This is small compared to the input capacitance of the preamplifiers of a few hundreds $\units{pF}$. Hence cross talk between neighbouring channels is expected to be below 5\%. In the data taken with Si sensors, no indication of cross talk is found.

\begin{figure}[ht!]
\begin{center}
    \includegraphics[width=0.65\columnwidth]{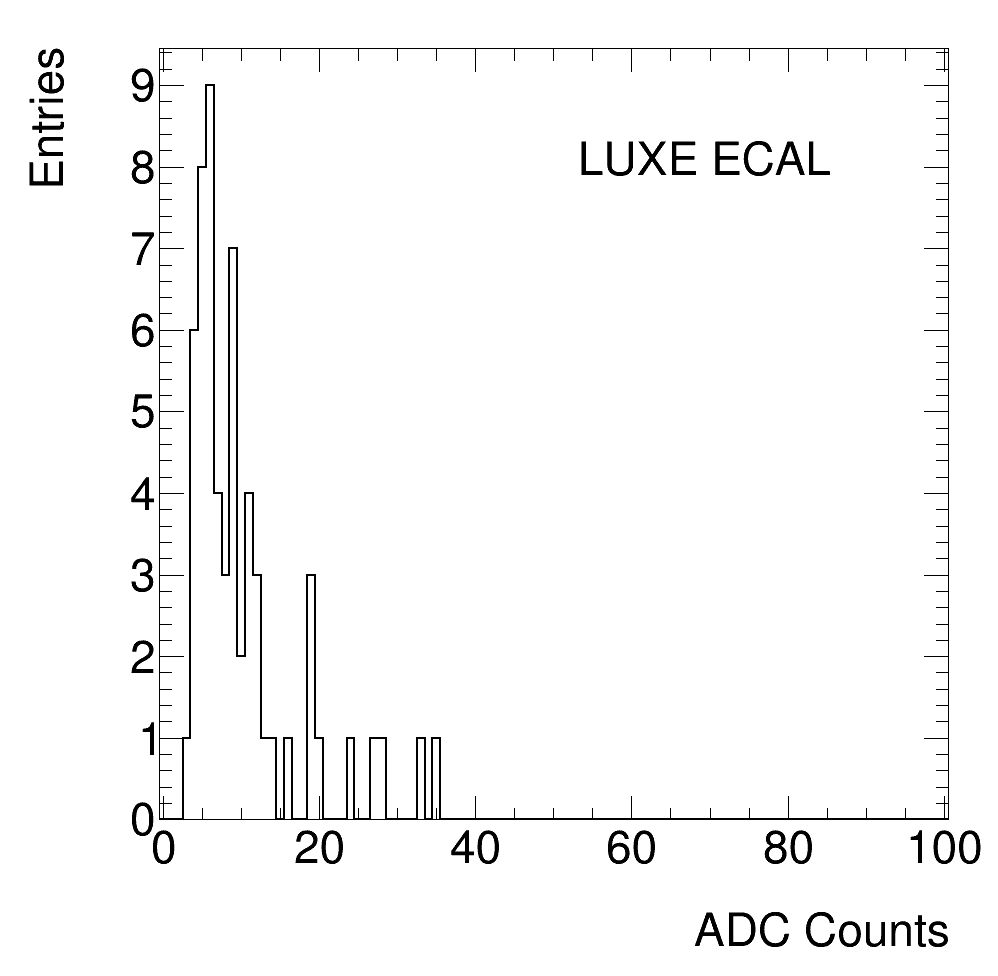}
    \caption{The distribution of signals from beam electrons hitting the area of the aluminium traces between two pads in the upper row on the GaAs sensor and assigned to pads to which the traces are connected.}
    \label{fig:signals_on_traces}
\end{center}
\end{figure}

The traces between pads of the GaAs sensor, as shown in Figs.~\ref{sensor_GAAS_pic} and~\ref{sensor_GAAS_cut}, are on the same potential as the pads. Beam electrons hitting the area of the traces release charge carriers in the sensor, and their drift in the electric field between traces and the back-plane may induce signals on the traces. These signals would be assigned to the pads to which the traces are connected, thus faking depositions far from e.g. a narrow shower of an electron. 
As seen in the picture of a GaAs sensor, shown in Fig.~\ref{sensor_GAAS_pic}, the bond pads on the top are connected to the sensor pads below via aluminium traces in the gaps between pad rows. The gaps in the upper row contain all strips. 
Using beam electrons with a trajectory pointing to a spot in the area of the traces between two chosen pads in the upper row, signals are searched for in pads connected to these traces, located in the rows below. Indeed such signals are found in pads below the upper row, and on the left side of the considered gap only i.e. the pads connected to the aluminium strips in this gap. The distribution of the registered signal size for a fixed number of beam electrons is shown in Fig.~\ref{fig:signals_on_traces}. As can be seen, the signals are, compared the the distribution in Fig.~\ref{fig:signal_silicon_GaAs}, relatively small. It should be noted though that only about 1\% of the electrons crossing the aluminium strip area, show a signal in the pads to which the strips are connected.
The integral of this distribution is a measure of the total signal size assigned to wrong pads. The ratio of this integral to the integral of the signals on the neighbour pad hit with the same beam intensity\footnote{A flat beam density over the pad and the trace area is assumed.} is about $2 \cdot 10^{-4}$. This is the worst case scenario, since the number of aluminium strips is decreasing when going down in the rows, and this fraction of wrongly assigned signal size is considered to be of tolerable impact for any calorimetric application. 

\section{Summary }\label{subsec8}

Two samples of Si pad sensors and two samples of GaAs sensors of 500\,\textmu m thickness and pad sizes of about 5$\times$5 mm$^2$ were studied in an electron beam of 5 GeV. These sensors are foreseen to be installed in a highly compact and segmented electromagnetic sandwich calorimeter, and will be positioned in thin gaps between tungsten plates. Therefore the signals from the pads are guided to the edges of the sensors where the front-end electronics is positioned. 
For the Si sensors, copper traces on a Kapton foil are used, connected to the sensor pads with conducting glue.
The pads of the GaAs sensors are connected to bond pads at the sensor edge via aluminium traces on the sensor substrate.

A full system test was performed. The sensors were read out by dedicated FE ASICs in 130\,nm CMOS technology, called FLAME. Data is taken with and without pre-processing and deconvolution performed with FPGAs. 

The signal distribution of single pads follows a Landau distribution convoluted with a Gaussian. A simulation of the expected signal distribution using \textsc{Geant4} is in very good agreement with data. The signal-to-noise ratio for single electrons amounts to 15 and 17 for Si and GaAs sensors, respectively. 

The response as a function of the local pad coordinate varies within a few per cent for the Si sensor. A larger drop is observed at the edges of the pads on the GaAs sensor. 
For Si sensors no signal loss between pads is observed. For GaAs sensors a loss of up to 40\% in the horizontal direction and up to 10\% in the vertical direction direction, in a range of $0.5 \units{mm}$ between adjacent pads is measured. The average response of the sensor pads varies between 3\% for Si sensors and 2\% for GaAs sensors.

For Si sensors no cross talk between pads and copper traces is seen. 
Signals due to hits in the trace region on the GaAs sensors, assigned to the pads connected to the traces, are observed but are small in size. Assuming a flat beam intensity, their integrated size, compared to the one of a pad, amounts to $2\cdot 10^{-4}$.

These technologies of very thin Si and GaAs sensor planes, readout via FLAME ASICs, are very promising for application in a highly compact and granular electromagnetic calorimeter. Traces on the substrate may potentially lead to smaller thickness of the sensor plane, however areas of signal loss between pads may require more effort in calibration and lead to a loss of precision in the shower position reconstruction.

\break

\bmhead*{Acknowledgments}

This research was supported by the 
European Union’s Horizon 2020 Research and Innovation programme 3309 under GA no. 101004761, the German-Israeli Foundation under grant no. I/1492/303.7/2019, the Israeli PAZY Foundation (ID 318), and the Generalitat Valenciana (Spain) under the grant number CIDEGENT/2020/21. IFIC members also acknowledge the financial support from the MCIN with funding from the European Union NextGenerationEU and Generalitat Valenciana in the call Programa de Planes Complementarios de I+D+i (PRTR 2022) through the project with reference ASFAE$/2022/015$; the Program \textit{Programa Estatal para Desarrollar, Atraer y Retener Talento}  PEICTI 2021-2023 through the project with reference CNS$2022-135420$; the Spanish MCIU/ AEI / 10.13039/501100011033 and European Union / FEDER via the grant PID2021-122134NB-C21 and 
the Generalitat Valenciana (GV) via the Excellence
Grant CIPROM/2021/073. Support was given by the
National Science Centre, Poland, under grant no. 2021/43/B/ST2/01107, 
and partially support by the Romanian Ministry of Research, Innovation and Digitalization under Romanian National Core Program LAPLAS VII - contract no. 30N/2023 and by the Executive Agency for Higher Education, Research, Development and Innovation, UEFISCDI, contract no. 9Sol(T9)/2024.

We would like to express our thanks to Dr. Anton Tyazhev from the Tomsk State University NI TSU, TSU, Tomsk, Russia for providing the GaAs sensors.

The measurements leading to these results have been performed at the Test Beam Facility at DESY Hamburg (Germany), a member of the Helmholtz Association (HGF).

\bibliography{sn-bibliography}%


\begin{thebibliography}{27}
\ifx \bisbn   \undefined \def \bisbn  #1{ISBN #1}\fi
\ifx \binits  \undefined \def \binits#1{#1}\fi
\ifx \bauthor  \undefined \def \bauthor#1{#1}\fi
\ifx \batitle  \undefined \def \batitle#1{#1}\fi
\ifx \bjtitle  \undefined \def \bjtitle#1{#1}\fi
\ifx \bvolume  \undefined \def \bvolume#1{\textbf{#1}}\fi
\ifx \byear  \undefined \def \byear#1{#1}\fi
\ifx \bissue  \undefined \def \bissue#1{#1}\fi
\ifx \bfpage  \undefined \def \bfpage#1{#1}\fi
\ifx \blpage  \undefined \def \blpage #1{#1}\fi
\ifx \burl  \undefined \def \burl#1{\textsf{#1}}\fi
\ifx \doiurl  \undefined \def \doiurl#1{\url{https://doi.org/#1}}\fi
\ifx \betal  \undefined \def \betal{\textit{et al.}}\fi
\ifx \binstitute  \undefined \def \binstitute#1{#1}\fi
\ifx \binstitutionaled  \undefined \def \binstitutionaled#1{#1}\fi
\ifx \bctitle  \undefined \def \bctitle#1{#1}\fi
\ifx \beditor  \undefined \def \beditor#1{#1}\fi
\ifx \bpublisher  \undefined \def \bpublisher#1{#1}\fi
\ifx \bbtitle  \undefined \def \bbtitle#1{#1}\fi
\ifx \bedition  \undefined \def \bedition#1{#1}\fi
\ifx \bseriesno  \undefined \def \bseriesno#1{#1}\fi
\ifx \blocation  \undefined \def \blocation#1{#1}\fi
\ifx \bsertitle  \undefined \def \bsertitle#1{#1}\fi
\ifx \bsnm \undefined \def \bsnm#1{#1}\fi
\ifx \bsuffix \undefined \def \bsuffix#1{#1}\fi
\ifx \bparticle \undefined \def \bparticle#1{#1}\fi
\ifx \barticle \undefined \def \barticle#1{#1}\fi
\bibcommenthead
\ifx \bconfdate \undefined \def \bconfdate #1{#1}\fi
\ifx \botherref \undefined \def \botherref #1{#1}\fi
\ifx \url \undefined \def \url#1{\textsf{#1}}\fi
\ifx \bchapter \undefined \def \bchapter#1{#1}\fi
\ifx \bbook \undefined \def \bbook#1{#1}\fi
\ifx \bcomment \undefined \def \bcomment#1{#1}\fi
\ifx \oauthor \undefined \def \oauthor#1{#1}\fi
\ifx \citeauthoryear \undefined \def \citeauthoryear#1{#1}\fi
\ifx \endbibitem  \undefined \def \endbibitem {}\fi
\ifx \bconflocation  \undefined \def \bconflocation#1{#1}\fi
\ifx \arxivurl  \undefined \def \arxivurl#1{\textsf{#1}}\fi
\csname PreBibitemsHook\endcsname

%%% 1
\bibitem{Moliere_1}
\begin{barticle}
\bauthor{\bsnm{Nelson}, \binits{W.R.}}, \betal:
\batitle{{Electron-Induced Cascade Showers in Copper and Lead at 1 GeV}}.
\bjtitle{Phys. Rev.}
\bvolume{149}(\bissue{1}),
\bfpage{201}
(\byear{1966}).
\doiurl{10.1103/physrev.149.201}
\end{barticle}
\endbibitem

%%% 2
\bibitem{Moliere_2}
\begin{barticle}
\bauthor{\bsnm{Bathow}, \binits{G.}}, \betal:
\batitle{{Measurements of the longitudinal and lateral development of electromagnetic cascades in lead, copper and aluminum at 6 GeV}}.
\bjtitle{Nucl. Phys. B}
\bvolume{20}(\bissue{3}),
\bfpage{592}
(\byear{1970}).
\doiurl{10.1016/0550-3213(70)90389-5}
\end{barticle}
\endbibitem

%%% 3
\bibitem{Abramowicz:2010bg}
\begin{barticle}
\bauthor{\bsnm{Abramowicz}, \binits{H.}}, \betal:
\batitle{{Forward Instrumentation for ILC Detectors}}.
\bjtitle{JINST}
\bvolume{5},
\bfpage{12002}
(\byear{2010})
{\href{https://arxiv.org/abs/1009.2433}{{arXiv:1009.2433}}}
{[physics.ins-det]}.
\doiurl{10.1088/1748-0221/5/12/P12002}
\end{barticle}
\endbibitem

%%% 4
\bibitem{LUXE:2023crk}
\begin{barticle}
\bauthor{\bsnm{Abramowicz}, \binits{H.}}, \betal:
\batitle{{Technical Design Report for the LUXE experiment}}.
\bjtitle{Eur. Phys. J. ST}
\bvolume{233}(\bissue{10}),
\bfpage{1709}--\blpage{1974}
(\byear{2024})
{\href{https://arxiv.org/abs/2308.00515}{{arXiv:2308.00515}}}
{[hep-ex]}.
\doiurl{10.1140/epjs/s11734-024-01164-9}
\end{barticle}
\endbibitem

%%% 5
\bibitem{Bhabha:1936zz}
\begin{barticle}
\bauthor{\bsnm{Bhabha}, \binits{H.J.}}:
\batitle{{The scattering of positrons by electrons with exchange on Dirac's theory of the positron}}.
\bjtitle{Proc. Roy. Soc. Lond. A}
\bvolume{154},
\bfpage{195}--\blpage{206}
(\byear{1936}).
\doiurl{10.1098/rspa.1936.0046}
\end{barticle}
\endbibitem

%%% 6
\bibitem{Afanaciev:2012im}
\begin{barticle}
\bauthor{\bsnm{Afanaciev}, \binits{K.}}, \betal:
\batitle{{Investigation of the radiation hardness of GaAs sensors in an electron beam}}.
\bjtitle{JINST}
\bvolume{7},
\bfpage{11022}
(\byear{2012}).
\doiurl{10.1088/1748-0221/7/11/P11022}
\end{barticle}
\endbibitem

%%% 7
\bibitem{Kruchonak:2020jsk}
\begin{barticle}
\bauthor{\bsnm{Kruchonak}, \binits{U.}}, \betal:
\batitle{{Radiation hardness of GaAs: Cr and Si sensors irradiated by electron beam}}.
\bjtitle{Nucl. Instrum. Meth. A}
\bvolume{975},
\bfpage{164204}
(\byear{2020})
{\href{https://arxiv.org/abs/2006.01254}{{arXiv:2006.01254}}}
{[physics.ins-det]}.
\doiurl{10.1016/j.nima.2020.164204}
\end{barticle}
\endbibitem

%%% 8
\bibitem{deBoer:2015dmy}
\begin{barticle}
\bauthor{\bparticle{de} \bsnm{Boer}, \binits{W.}}, \betal:
\batitle{{A fourfold segmented silicon strip sensor with read-out at the edges}}.
\bjtitle{Nucl. Instrum. Meth. A}
\bvolume{788},
\bfpage{154}--\blpage{160}
(\byear{2015}).
\doiurl{10.1016/j.nima.2015.03.082}
\end{barticle}
\endbibitem

%%% 9
\bibitem{Abramowicz:2018vwb}
\begin{barticle}
\bauthor{\bsnm{Abramowicz}, \binits{H.}}, \betal:
\batitle{{Performance and Moli\`ere radius measurements using a compact prototype of LumiCal in an electron test beam}}.
\bjtitle{Eur. Phys. J. C}
\bvolume{79}(\bissue{7}),
\bfpage{579}
(\byear{2019})
{\href{https://arxiv.org/abs/1812.11426}{{arXiv:1812.11426}}}
{[physics.ins-det]}.
\doiurl{10.1140/epjc/s10052-019-7077-9}
\end{barticle}
\endbibitem

%%% 10
\bibitem{Breit:1934zz}
\begin{barticle}
\bauthor{\bsnm{Breit}, \binits{G.}},
\bauthor{\bsnm{Wheeler}, \binits{J.A.}}:
\batitle{{Collision of two light quanta}}.
\bjtitle{Phys. Rev.}
\bvolume{46}(\bissue{12}),
\bfpage{1087}--\blpage{1091}
(\byear{1934}).
\doiurl{10.1103/PhysRev.46.1087}
\end{barticle}
\endbibitem

%%% 11
\bibitem{tomita2014study}
\begin{botherref}
\oauthor{\bsnm{Tatsuhiko}, \binits{T.}}, et al.:
{A study of silicon sensor for ILD ECAL}
(2014)
{\href{https://arxiv.org/abs/1403.7953}{{arXiv:1403.7953}}}
{[physics.ins-det]}
\end{botherref}
\endbibitem

%%% 12
\bibitem{tyazhev2021}
\begin{botherref}
\oauthor{\bsnm{Tyazhev}, \binits{A.}}:
Novel highly compact electromagnetic calorimeters based on High Resistive GaAs:Cr sensors.
IEEE NUCLEAR SCIENCE SYMPOSIUM AND MEDICAL IMAGING CONFERENCE
(2021)
\end{botherref}
\endbibitem

%%% 13
\bibitem{FLAME_1}
\begin{botherref}
\oauthor{\bsnm{Moron}, \binits{J.}}:
FLAME SoC readout ASIC for electromagnetic calorimeter.
\url{https://indico.cern.ch/event/1127562/contributions/4904506/attachments/2512388/4318796/moron_TWEPP_2022_09_21.pdf}.
TWEPP 2022 Topical Workshop on Electronics for Particle Physics, Bergen, Norway
(September 19-23, 2022)
\end{botherref}
\endbibitem

%%% 14
\bibitem{FLAME_2}
\begin{botherref}
\oauthor{\bsnm{Idzik}, \binits{M.}}:
The FLAME and FLAXE ASICs.
\url{https://agenda.infn.it/event/36206/contributions/202659/attachments/106949/150868/idzik_FEE_2023_06_FLAME.pdf}.
XII Front-end Electronics Workshop, Torino, Italy
(2023)
\end{botherref}
\endbibitem

%%% 15
\bibitem{Kulis:2011zz}
\begin{barticle}
\bauthor{\bsnm{Kulis}, \binits{S.}},
\bauthor{\bsnm{Idzik}, \binits{M.}}:
\batitle{{Triggerless readout with time and amplitude reconstruction of event based on deconvolution algorithm}}.
\bjtitle{Acta Phys. Polon. Supp.}
\bvolume{4},
\bfpage{49}--\blpage{58}
(\byear{2011}).
\doiurl{10.5506/APhysPolBSupp.4.49}
\end{barticle}
\endbibitem

%%% 16
\bibitem{test-beam}
\begin{barticle}
\bauthor{\bsnm{Diener}, \binits{R.}}, \betal:
\batitle{{The DESY II test beam facility}}.
\bjtitle{Nucl. Instr. and Meth.}
\bvolume{922},
\bfpage{265}--\blpage{286}
(\byear{2019}).
\doiurl{10.1016/j.nima.2018.11.133}
\end{barticle}
\endbibitem

%%% 17
\bibitem{Baesso:2019smg}
\begin{barticle}
\bauthor{\bsnm{Baesso}, \binits{P.}}, \betal:
\batitle{{The AIDA-2020 TLU: a flexible trigger logic unit for test beam facilities}}.
\bjtitle{JINST}
\bvolume{14}(\bissue{09}),
\bfpage{09019}
(\byear{2019})
{\href{https://arxiv.org/abs/2005.00310}{{arXiv:2005.00310}}}
{[physics.ins-det]}.
\doiurl{10.1088/1748-0221/14/09/P09019}
\end{barticle}
\endbibitem

%%% 18
\bibitem{Ahlburg:2019jyj}
\begin{barticle}
\bauthor{\bsnm{Ahlburg}, \binits{P.}}, \betal:
\batitle{{EUDAQ-a data acquisition software framework for common beam telescopes}}.
\bjtitle{JINST}
\bvolume{15}(\bissue{01}),
\bfpage{01038}
(\byear{2020})
{\href{https://arxiv.org/abs/1909.13725}{{arXiv:1909.13725}}}
{[physics.ins-det]}.
\doiurl{10.1088/1748-0221/15/01/P01038}
\end{barticle}
\endbibitem

%%% 19
\bibitem{Liu:2019wim}
\begin{barticle}
\bauthor{\bsnm{Liu}, \binits{Y.}}, \betal:
\batitle{{EUDAQ2\textemdash{}A flexible data acquisition software framework for common test beams}}.
\bjtitle{JINST}
\bvolume{14}(\bissue{10}),
\bfpage{10033}
(\byear{2019})
{\href{https://arxiv.org/abs/1907.10600}{{arXiv:1907.10600}}}
{[physics.ins-det]}.
\doiurl{10.1088/1748-0221/14/10/P10033}
\end{barticle}
\endbibitem

%%% 20
\bibitem{Dannheim:2020jlk}
\begin{barticle}
\bauthor{\bsnm{Dannheim}, \binits{D.}}, \betal:
\batitle{{Corryvreckan: A Modular 4D Track Reconstruction and Analysis Software for Test Beam Data}}.
\bjtitle{JINST}
\bvolume{16}(\bissue{03}),
\bfpage{03008}
(\byear{2021})
{\href{https://arxiv.org/abs/2011.12730}{{arXiv:2011.12730}}}
{[physics.ins-det]}.
\doiurl{10.1088/1748-0221/16/03/P03008}
\end{barticle}
\endbibitem

%%% 21
\bibitem{Duda:1972ymn}
\begin{barticle}
\bauthor{\bsnm{Duda}, \binits{R.O.}},
\bauthor{\bsnm{Hart}, \binits{P.E.}}:
\batitle{{Use of the Hough transformation to detect lines and curves in pictures}}.
\bjtitle{Commun. ACM}
\bvolume{15}(\bissue{1}),
\bfpage{11}
(\byear{1972}).
\doiurl{10.1145/361237.361242}
\end{barticle}
\endbibitem

%%% 22
\bibitem{1990PaReL11167B}
\begin{barticle}
\bauthor{\bsnm{{Ben-Tzvi}}, \binits{D.}},
\bauthor{\bsnm{{Sandler}}, \binits{M.B.}}:
\batitle{{A combinatorial Hough transform}}.
\bjtitle{Pattern Recognition Letters}
\bvolume{11}(\bissue{3}),
\bfpage{167}--\blpage{174}
(\byear{1990}).
\doiurl{10.1016/0167-8655(90)90002-J}
\end{barticle}
\endbibitem

%%% 23
\bibitem{AGOSTINELLI2003250}
\begin{barticle}
\bauthor{\bsnm{Agostinelli}, \binits{S.}}, \betal:
\batitle{Geant4 - a simulation toolkit}.
\bjtitle{Nucl. Instrum. Meth. A}
\bvolume{506}(\bissue{3}),
\bfpage{250}--\blpage{303}
(\byear{2003}).
\doiurl{10.1016/S0168-9002(03)01368-8}
\end{barticle}
\endbibitem

%%% 24
\bibitem{BRUN199781}
\begin{barticle}
\bauthor{\bsnm{Brun}, \binits{R.}},
\bauthor{\bsnm{Rademakers}, \binits{F.}}:
\batitle{Root — an object oriented data analysis framework}.
\bjtitle{Nucl. Instrum. Meth. A}
\bvolume{389}(\bissue{1}),
\bfpage{81}--\blpage{86}
(\byear{1997}).
\doiurl{10.1016/S0168-9002(97)00048-X}
\end{barticle}
\endbibitem

%%% 25
\bibitem{Fabjan:2020wnt}
\begin{bbook}
\beditor{\bsnm{Fabjan}, \binits{C.W.}},
\beditor{\bsnm{Schopper}, \binits{H.}} (eds.):
\bbtitle{{P}article {P}hysics {R}eference {L}ibrary: {V}olume 2: {D}etectors for {P}articles and {R}adiation}.
\bpublisher{Springer},
\blocation{Cham}
(\byear{2020}).
\doiurl{10.1007/978-3-030-35318-6}
\end{bbook}
\endbibitem

%%% 26
\bibitem{PhysRevB.13.761}
\begin{barticle}
\bauthor{\bsnm{Nam}, \binits{S.B.}},
\bauthor{\bsnm{Reynolds}, \binits{D.C.}},
\bauthor{\bsnm{Litton}, \binits{C.W.}},
\bauthor{\bsnm{Almassy}, \binits{R.J.}},
\bauthor{\bsnm{Collins}, \binits{T.C.}},
\bauthor{\bsnm{Wolfe}, \binits{C.M.}}:
\batitle{{Free-exciton energy spectrum in GaAs}}.
\bjtitle{Phys. Rev. B}
\bvolume{13},
\bfpage{761}--\blpage{767}
(\byear{1976}).
\doiurl{10.1103/PhysRevB.13.761}
\end{barticle}
\endbibitem

%%% 27
\bibitem{AYZENSHTAT2002120}
\begin{barticle}
\bauthor{\bsnm{{Ayzenshtat A.I. and others}}}:
\batitle{{GaAs as a material for particle detectors}}.
\bjtitle{Nuclear Instruments and Methods in Physics Research Section A: Accelerators, Spectrometers, Detectors and Associated Equipment}
\bvolume{494}(\bissue{1}),
\bfpage{120}--\blpage{127}
(\byear{2002}).
\doiurl{10.1016/S0168-9002(02)01455-9}.
\bcomment{Proceedings of the 8th International Conference on Instrumentation for Colliding Beam Physics}
\end{barticle}
\endbibitem

\end{thebibliography}

\end{document}